\documentclass[11pt,preprint]{aastex}

\newcommand{\halpha}{H$\alpha$}
\newcommand{\hi}{H\thinspace I}
\newcommand{\hii}{H\thinspace II}
\newcommand{\hei}{He\thinspace I}
\newcommand{\heii}{He\thinspace II}
\newcommand{\cii}{C\thinspace II}
\newcommand{\nii}{N\thinspace II}
\newcommand{\niii}{N\thinspace III}
\newcommand{\oii}{O\thinspace II}
\newcommand{\oiii}{O\thinspace III}
\newcommand{\sii}{S\thinspace II}
\newcommand{\siii}{S\thinspace III}
\newcommand{\neii}{Ne\thinspace II}
\newcommand{\neiii}{Ne\thinspace III}

\shorttitle{COMPOSITION GRADIENT IN M101. II.}
\shortauthors{KENNICUTT, BRESOLIN \& GARNETT}

\begin{document}


\title{The Composition Gradient in M101 Revisited. II.  Electron
Temperatures and Implications for the Nebular Abundance Scale} 

\medskip
\author{Robert C.\ Kennicutt, Jr.\altaffilmark{1}}
\affil{Steward Observatory, University of Arizona, Tucson, AZ  85721}
\email{rkennicutt@as.arizona.edu}

\author{Fabio Bresolin}
\affil{Institute for Astronomy, 2680 Woodlawn Drive, Honolulu, HI  96822}
\email{bresolin@ifa.hawaii.edu}

\and

\author{Donald R. Garnett}
\affil{Steward Observatory, University of Arizona, Tucson, AZ  85721}
\email{dgarnett@as.arizona.edu}

\altaffiltext{1}{Observations reported here were obtained at 
the MMT Observatory, a joint facility of the
University of Arizona and the Smithsonian Institution.} 

\begin{abstract}

We use high signal/noise spectra of 20 \hii\ regions in the giant
spiral galaxy M101 to derive electron temperatures for the \hii\
regions and robust metal abundances over radii R = 0.19--1.25 R$_0$
(6--41 kpc).  We compare the consistency of electron temperatures
measured from the [O~III]$\lambda$4363, [N~II]$\lambda$5755,
[S~III]$\lambda$6312, and [O~II]$\lambda$7325 auroral
lines. Temperatures from [O~III], [S~III], and [N~II] are correlated
with relative offsets that are consistent with expectations from
nebular photoionization models. However, the temperatures derived from
the [O~II]$\lambda$7325 line show a large scatter and are nearly
uncorrelated with temperatures derived from other ions. We tentatively
attribute this result to observational and physical effects, which may
introduce large random and systematic errors into abundances derived
solely from [O~II] temperatures. Our derived oxygen abundances are
well fitted by an exponential distribution over six disk scale lengths, 
from approximately
1.3\,(O/H)$_\odot$ in the center to 1/15\,(O/H)$_\odot$ in the
outermost region studied (for solar 12 + log\,(O/H)~=~8.7). We measure
significant radial gradients in N/O and He/H abundance ratios, but
relatively constant S/O and Ar/O.  Our results are in approximate
agreement with previously published abundances studies of M101 based
on temperature measurements of a few \hii\ regions.  However our
abundances are systematically lower by 0.2--0.5 dex than those derived
from the most widely used strong-line ``empirical" abundance
indicators, again consistent with previous studies based on smaller
\hii\ region samples.  Independent measurements of the Galactic
interstellar oxygen abundance from ultraviolet absorption lines are in
good agreement with the $T_e$-based nebular abundances.  We suspect
that most of the disagreement with the strong-line abundances arises
from uncertainties in the nebular models that are used to calibrate
the ``empirical" scale, and that strong-line abundances derived for
\hii\ regions and emission-line galaxies are as much as a factor of
two higher than the actual oxygen abundances.  However other
explanations, such as the effects of temperature fluctuations on the
auroral line based abundances cannot be completely ruled out.  These
results point to the need for direct abundance determinations of a
larger sample of extragalactic \hii\ regions, especially for objects
more metal-rich than solar.

\end{abstract}


\section{INTRODUCTION}

The spiral galaxy M101 (NGC 5457) has played a central role in 
our understanding of gas-phase abundance distributions in galaxies. Its 
large disk ($R_0$ = 14.4 arcmin = 32.4 kpc; de Vaucouleurs et al.~1991) 
contains well over 1000 catalogued \hii\ regions (Hodge et al.~1990;
Scowen, Dufour, \& Hester 1992) with a large radial range in excitations
and metal abundances.  Spectrophotometry of its brightest \hii\ regions
by Searle (1971) and Smith (1975) provided some of the first quantitative
measurements of radial abundance gradients in disks.

The first systematic studies of \hii\ region abundances in galaxies
were based on ``direct" methods, in which measurements of 
temperature-sensitive auroral lines (usually [\oiii]$\lambda$4363)
were used to constrain the temperature and emissivities of the 
abundance-sensitive forbidden lines directly.  M101 was the object
of several such studies, each based on electron temperature ($T_e$)
measurements of 1--5 bright \hii\ regions (Searle 1971; Smith 1975; 
Shields \& Searle 1978; Rosa 1981; Sedwick \& Aller 1981; 
Rayo, Peimbert, \& Torres-Peimbert 1982; Torres-Peimbert, 
Peimbert, \& Fierro 1989; Kinkel \& Rosa 1994; Garnett \& Kennicutt 1994).  
These studies established the existence of a roughly exponential abundance
gradient in the disk, and with roughly a fivefold range in 
abundance over the range of radii studied ($\sim$0.2--0.8 $R_0$).
They also established that the
pronounced nebular excitation variations observed among extragalactic \hii\
regions are driven primarily by changes in their electron temperatures.
For compositions typical of present-day disks
the dominant ionized gas coolants are fine-structure lines of heavy
elements, so the average electron temperature of an \hii\ region 
decreases with increasing metal abundance.

The application of this ``direct" abundance technique has been hampered 
by several technical limitations, however. 
For oxygen abundances much above $\sim$0.5\,(O/H)$_\odot$, increased
cooling by IR fine-structure lines causes the [\oiii]$\lambda$4363 
line to become increasingly weak and difficult to detect, rendering
direct abundance determinations based on this feature impossible.
Questions about the linearity of photon-counting detectors used prior
to the 1990s also reduced confidence in the abundances even when the 
[\oiii]$\lambda$4363 line was detected (e.g., Torres-Peimbert et al.~1989).
Moreover, the [\oiii] temperature itself may not be a representative
measure of the average temperature in regions where O$^{++}$ occupies 
a small fraction of the nebular volume or have significant internal
gradients in $T_e$ (e.g., Stasi\'nska 1978; Garnett 1992; Oey \& Kennicutt 
1993). Local fluctuations in electron temperature may also affect the
temperature derived from integrated measurements of collisionally
excited lines in the optical part of the spectrum (e.g., Peimbert 1967).

The absence of direct abundance measurements for metal-rich objects
led to the development of ``empirical" methods, in which the ratios of
strong lines ([\oii]$\lambda$3727, [\oiii]$\lambda\lambda$4959, 5007,
[\nii]$\lambda$6583, [\sii]$\lambda\lambda$6717,6731,
[\siii]$\lambda\lambda$9069,9532) are used to estimate the oxygen
abundance, without any direct electron temperature measurement (e.g.,
Pagel et al.~1979; Alloin et al.~1979; Edmunds \& Pagel 1984; 
Dopita \& Evans 1986; McGaugh 1991; 
V\'{\i}lchez \& Esteban 1996; van Zee et al.~1998; Dutil \& Roy 1999; 
D\'{\i}az \& P\'erez-Montero 2000; Kewley \& Dopita 2002).  These methods rely
on the physical coupling betweeen metal abundance and electron
temperature, coupled with the observation of relatively tight
correlations of forbidden-line ratios among extragalactic \hii\
regions (e.g., Baldwin, Phillips, \& Terlevich 1981; McCall, Rybski,
\& Shields 1985).  Early applications of these empirical methods were
calibrated using a combination of $T_e$-based abundances for
metal-poor objects and nebular photoionization models at higher
abundances (see Garnett 2003 for a review).  However, most subsequent
calibrations have relied increasingly on grids of \hii\ region
photoionization models (e.g., McGaugh 1991; Kewley \& Dopita 2002).

Over the past 15--20 years the overwhelming majority of published
abundance studies of spiral galaxies have been based on these
strong-line methods.  These have included several comprehensive
studies of the abundance trends across large galaxy samples (e.g.,
McCall et al.~1985; Skillman, Kennicutt, \& Hodge 1989; Vila-Costas \&
Edmunds 1992; Oey \& Kennicutt 1993, Martin \& Roy 1994; Zaritsky,
Kennicutt, \& Huchra 1994; Skillman et al.\ 1996; Pilyugin et
al.\ 2002), and applications to the integrated spectra of high-redshift
galaxies (e.g., Kobulnicky \& Zaritsky 1999; Carollo \& Lilly 2001;
Kobulnicky et al.~2003).  These methods have been
applied to M101 by Scowen et al.~(1992), Kennicutt \& Garnett (1996,
hereafter Paper~I) and Pilyugin (2001b).

The widespread application of these ``empirical" abundances has hardly
lessened the need for a foundation of abundance measurements based on
high-quality, robust electron-temperature measurements.  Fortunately,
the introduction of high-efficiency spectrographs with CCD detectors
on 4--10 m class telescopes now makes such an undertaking possible,
through measurements of other auroral lines such as
[\siii]$\lambda$6312, [\nii]$\lambda$5755, and
[\oii]$\lambda\lambda$7320,7330 (e.g., Kinkel \& Rosa 1994,
Castellanos, D\'\i az, \& Terlevich 2002 and references therein).  The
[\siii]$\lambda$6312 line is an especially attractive thermometer,
because it is observable to higher metal abundances than
[\oiii]$\lambda$4363, and often samples a larger fraction of the
nebular volume in these regions.  Its chief disadvantage is the need
for accurate spectrophotometry of the corresponding nebular lines of
[\siii]$\lambda\lambda$9069,9532 in the near-infrared, but with the
advent of red-sensitive CCD detectors measurements of these lines are
feasible (e.g., Garnett 1989; Bresolin, Kennicutt, \& Garnett 1999
[hereafter BKG]; Castellanos et al.~2002).

This paper is the second in a series on the spectral properties and
chemical abundances in M101.  In Paper~I 
we presented strong-line measurements of [\oii], [\oiii], [\nii],
[SII], and [\siii] for 41 \hii\ regions, and used several empirical
abundance calibrations to examine the radial and azimuthal form of
the oxygen abundance distribution.  In this paper we use higher
quality spectra for 26 of these \hii\ regions obtained with the MMT telescope 
to measure electron temperatures and temperature-based abundances.
Our study is motivated by several goals:  

i) To enlarge the sample of M101 \hii\ regions with well-measured direct 
abundances (from 6 to 20), thereby enabling
us to separate genuine radial trends in the abundances from local 
variations or other spurious effects.  

ii) To extend by more than 50\% the radial coverage of the 
abundance measurements, to 0.19--1.25 $R_0$.

iii) To take advantage of the large homogeneous data set and the large
range of oxygen abundances in M101, to test for radial trends in N/O,
S/O, Ar/O, and He/H abundance ratios, with improved sensitivity over
most previous studies.

iv) To obtain robust CCD-based [\oiii] temperature measurements,
without the systematic uncertainties introduced by
detector nonlinearities in most of the previously published spectra.

v) To measure electron temperatures for two or more ions in a majority
of the \hii\ regions, which enables us to compare the temperature
scales, test the reliability of the individual auroral lines, and test
the relations between ionic temperatures that have been predicted by
nebular models.

vi) To compare our temperature-based abundances 
with the published ``empirical" strong-line calibrations, and with 
independent abundance constraints,
in order to test the consistency of the respective scales and 
understand the physical origins of any 
systematic differences.


The remainder of this paper is organized as follows.  In \S~2 we
describe the observations and data analysis procedures, and present
the spectrophotometric data used in this analysis.  In \S~3 we use the
data to derive electron temperatures and O, N, S, and Ne abundances,
and discuss the consistency of the temperature scales.  We analyze the
radial abundance distributions in M101 themselves in \S~4, and in \S~5
we combine our new data with high-quality direct abundances from the
literature, to compare the consistency of the direct and strong-line
abundance scales.  We conclude with a general discussion of the zeropoint
and reliability of the extragalactic nebular abundance scale 
in \S~6.  To maintain consistency with Paper~I we will adopt a distance
to M101 of 7.5 Mpc (Kelson et al.~1995) throughout this paper.
However, we now adopt the new solar oxygen abundance scale (12 +
log\,(O/H)$_\odot$~=~8.70) following Allende Prieto et al.~(2001) and
Holweger (2001).

\section{DATA}

Table 1 lists the objects measured in this study.  Following Paper~I
we have used the designations of Hodge et al.\ (1990) when available.
Detailed information on the positions and alternate designations of
these \hii\ regions can be found in Paper~I, and are not repeated here.

Our sample was selected to cover the maximum range in radius and 
abundance possible. 
For the outer disk ($R > 0.3~R_0$) we measured most of the bright knots 
in the large complexes NGC 5447, 5455, 5461, 5462, and 5471, 
as well as other, isolated 
\hii\ regions at intermediate radii.  
We also made an effort to obtain good azimuthal coverage, 
to avoid any bias that might be introduced by the possible asymmetry 
in abundance distributions in M101 (Paper~I).  Since we were aiming to 
detect at least one auroral line with high signal/noise, we concentrated 
on the brightest \hii\ regions at any radius.

The metal-rich \hii\ regions in the inner disk of M101 presented 
particular challenges.  The low excitation of the \hii\ regions
in the inner disk makes detection of the auroral lines difficult,
and the problem is exacerbated by the stronger stellar continua
in most regions (i.e., lower emission-line equivalent widths;
see Paper~I) and the relative absence of bright \hii\ regions.  We 
obtained multiple integrations for 4 \hii\ regions with $R < 0.3~R_0$
(H972, H974, H1013, H336), but only the latter two objects yielded
useful measurements of temperature-sensitive lines (\S~3.1).  

In addition to \hii\ regions from Paper~I, we also observed the
outermost \hii\ region in the survey of Scowen et al.~(1992) at 
(J2000) RA = 14$^h$ 03$^m$ 50\fs09, Dec = 54$^\circ$ 38\arcmin\ 05\farcs0, 
at a radial distance of 18\arcmin.0 (41 kpc) from the nucleus of M101.  
It is designated as SDH323 in Table 1 and the remainder of this paper.
We also include in our analysis measurements of the outer disk
\hii\ region H681 from Garnett \& Kennicutt (1994) made with the
MMT Red Channel Spectrograph, and analyzed using the same procedures
as described below.

\subsection{Spectrophotometry}

Most of the spectroscopic data used in this paper were obtained in
1994--1997 with the Blue Channel Spectrograph on the 4.5~m MMT Telescope.
An additional
spectrum of one object (H336 = Searle~5) was obtained in 2002 January
using the same spectrograph on the upgraded 6.5~m MMT.  In addition,
fluxes of lines longward of 6800 \AA\ were taken from data obtained
on the MMT with the Red Channel Spectrograph, as described in Paper~I.

Spectra covering the region 3650 -- 7000 \AA\
were obtained using a 500 grooves mm$^{-1}$ grating blazed at 5410 \AA,
used in first order with a 3600 \AA\ cutoff blocking filter.
A 2\arcsec $\times$ 180\arcsec\ slit yielded a spectral resolution
of 6 \AA\ FWHM.  The spectrograph was equipped with a 3072 $\times$ 1024
element thinned Loral CCD detector, providing a spatial scale of 
$0\farcs6$ per (binned) pixel.  \hii\ regions were observed for 1--3
integrations of 1200 s each.  This provided high signal/noise 
(S/N $\gg$ 50) measurements of the principle diagnostic emission lines 
and measurements of faint lines including [\oiii]$\lambda$4363 and
[\siii]$\lambda$6312 in all but the lowest-excitation \hii\ regions.

For the low-excitation metal-rich \hii\ regions, higher-resolution
spectra were obtained to provide more sensitivity to the faint
temperature-sensitive [\siii]$\lambda$6312 and
[\oii]$\lambda\lambda$7320,7330 lines.  H336 (Searle~5), H409
(NGC~5455), H1013, and H1105 (NGC~5461) were observed using a 1200
grooves mm$^{-1}$ grating blazed at 4830 \AA, tuned to cover the
wavelength range 6000 -- 7500 \AA\ with a resolution of 2.4 \AA\ FWHM.
Integration times ranged from 600 s for the NGC regions to 2400 s for
the fainter and more metal-rich H336 and H1013.  The flux scale of
these spectra were tied to the full-coverage spectra described above
at H$\alpha$.

For spectral lines longward of 7000 \AA\ (mainly [\oii]$\lambda$7320,7330
and [\siii]$\lambda$9069,9532) we used echellette spectra obtained with
the Red Channel Spectrograph on the MMT, as described in detail in 
Paper~I.  These spectra cover the wavelength range 4700 -- 10000 \AA,
and were taken with a 2\arcsec $\times$ 20\arcsec\ slit, yielding 
a spectral resolution of 7 $-$ 15 \AA\ FWHM.  The flux
scales of the two sets of spectra were tied together at H$\alpha$.

Standard stars from the list of Massey et al.~(1988) were observed
several times per night to determine the flux calibration.  Care
was taken to orient the spectrograph slit with the atmospheric
dispersion direction for all standard star measurements, and 
for the \hii\ regions when the dispersion was significant.  
Standard dark, flatfield (including twilight sky measurements),
and HeNeAr lamp exposures were taken to calibrate CCD sensitivity
variations, camera distortions, and the wavelength zeropoint.

The spectra were reduced using the TWODSPEC package in the NOAO IRAF
package\footnote{IRAF is distributed by the National Optical Astronomy 
Observatories, which are operated by the Association of Universities for 
Research in Astronomy, Inc., under cooperative agreement with the National
Science Foundation.}, using standard procedures.  Since more than one
object sometimes was observed along the slit, the spectra were fully
reduced and calibrated in two dimensions before extracting one-dimensional
flux-calibrated spectra for each \hii\ region.  The size of the extraction
aperture was adjusted depending on the size of the \hii\ region (typically
6\arcsec\ -- 10\arcsec), and was matched to the aperture used in the Red 
Channel spectra from Paper~I.

Line intensities were measured by integration of the flux under the
line profiles between two continuum points identified on each side of
the line. For partly overlapping lines (e.g.,
H$\alpha$+[\nii]$\lambda\lambda6563,6548,6583$ and
[SII]$\lambda\lambda6716,6731$) gaussian profiles have been assumed to
deblend the different components. Average spectra were measured in
those cases where multiple observations with the same grating were
taken. When available, line fluxes from the higher resolution spectra
were retained for the subsequent analysis, after a consistency check
with the lower resolution spectra was carried out.

The accuracy of the spectrophotometry was checked in several ways,
from consistency of the standard star calibrations on a given run
(typically $\pm$2--3\% rms over the entire wavelength range),
comparison of multiple measurements of the same \hii\ regions, and, in
the case of the faint lines, by Poisson statistics.  Uncertainties for
individual line measurements are listed in Table~1.

The observed spectra were corrected for interstellar reddening using 
the Balmer emission line decrement and the interstellar reddening law
of Cardelli, Clayton, \& Mathis (1989). We corrected for underlying
stellar Balmer absorption by determining iteratively the equivalent 
width of the underlying stellar absorption that provides consistent 
values of the extinction $c(H\beta)$ for all the lines considered 
(H$\alpha$ through H$\delta$). For the intrinsic Balmer line ratios 
we have assumed the theoretical Case B values of Hummer \& Storey 
(1987) at the temperature determined from the auroral lines.

\section{ELECTRON TEMPERATURES AND ABUNDANCES}

Analysis of the spectra was performed using the five-level atom
program of Shaw \& Dufour (1995), as implemented in the IRAF package
with the exception of [\siii], where we used updated collision strengths
from Tayal \& Gupta (1999). These yield [\siii] temperatures that are
about 200~K cooler and S$^{+2}$ abundances that are approximately 
0.1 dex higher than produced by the 5-level program.
Inspection of the [SII]$\lambda\lambda$6717,6731
doublet ratios in our \hii\ regions shows that most of the emitting
gas is at low electron densities ($n_e \ll 100$ cm$^{-3}$), so
collisional effects on the derived temperatures and abundances should
be negligible for {\it most} of the measured lines (possible
exceptions are discussed below).

\subsection{Electron Temperatures}

Electron temperatures were derived using four sets of auroral/nebular
line intensity ratios:
[\oiii]$\lambda$4363/[\oiii]$\lambda\lambda$4959,5007,
[\siii]$\lambda$6312/[\siii]$\lambda\lambda$9069,9532,
[\nii]$\lambda$5755/[\nii]$\lambda\lambda$6548,6583, and
[\oii]$\lambda\lambda$7319,7330/[\oii]$\lambda\lambda$3726,3729.  In
many of the \hii\ regions one or more of the auroral lines was too
faint to yield a reliable temperature measurement.  Care must be taken
to exclude low signal/noise auroral line measurements, to avoid having
the temperatures biased by incompleteness effects.  We propagated the
measured line flux errors directly into the $T_e$ determination, and
in the subsequent abundance analysis we discarded any data for which
$\sigma(T_e) \ge 1000~K$ or 10\%\ (whichever was larger).  This left
20 \hii\ regions with a $T_e$ measurement from at least one auroral
line, 17 objects with at least two temperature measurements, and 14
regions with three measurements. Three regions originally in the
sample, H203, H972 and H974, did not yield any reliable auroral line
detection, and these objects were dropped from the remaining analysis.
Table~2 lists the derived electron temperatures and uncertainties for
the sample.  For most objects the uncertainties in the $T_e$ values
are in the range $\pm$200--700 K.

The availability of multiple $T_e$ measurements for a large sample
of \hii\ regions allows us to compare the consistency of the various
ionic temperature scales.  Photoionization models can be used to
predict scaling relations between temperatures measured in different
ions, and these are commonly used in abundance analyses when only
a single temperature measurement (usually [\oiii]$\lambda$4363)
is available.  Garnett (1992) carried out an extensive analysis of
ion-weighted temperatures over the abundance range of interest, and
we can compare the actual temperature relationships with those
predicted by his models.

Figure~1 compares the measured [\oii], [\siii], and [\nii] temperatures
with [\oiii] temperatures derived for the M101 \hii\ regions in Table 2
(solid points).  In order to enlarge the comparison and check the
reliability of our results we also include data from a study of NGC~2403
by Garnett et al.~(1997).
These latter data are plotted as open points.  
The dashed lines show the scaling relations predicted by the photoionization
models of Garnett (1992):

\begin{equation}
\rm T[\siii] = 0.83\ T[\oiii] + 1700~K
\end{equation}

\begin{equation}
\rm T[\nii] = T[\oii] = 0.70\ T[\oiii] + 3000~K
\end{equation}  

The best correlation is between [\oiii] and [\siii] temperatures (T[\oiii]
and T[\siii], respectively), as shown in the top 
panel of Figure~1.  Apart from a few objects with large uncertainties
in one or both auroral line measurements, the respective temperatures
roughly follow the theoretically predicted relationship from Garnett (1992).
A similar correlation was observed by Vermeij \& van der Hulst (2002)
for a sample of Magellanic Cloud \hii\ regions, albeit with larger
observational uncertainties than in this study.  
This correlation is reassuring, and may alleviate some of the concerns
raised about the utility of [\oiii] temperatures in moderate-abundance 
\hii\ regions (e.g., Stasi\'nska 2002).  Despite the different ionization
thresholds for S$^{++}$ and O$^{++}$ (23 eV vs 35 eV, respectively), 
the ions yield consistent measurements of $T_e$
over the range of temperatures where both auroral lines can be observed.
In particular there is no evidence for [\oiii] temperatures being
systematically higher in the higher-abundance, lower-temperature \hii\
regions, as might be expected from temperature fluctuation effects.

The comparison of [\nii] and [\oiii] temperatures (middle panel in Fig.~1)
is less clear, because
only a few regions have measured [\nii]$\lambda$5755 lines, and for most
of them the T[\nii] is uncertain ($\sigma(T_e) > 1000~K$).  
The four regions with the
best-determined temperatures show good agreement with the expectations
from the Garnett (1992) models, but more data are needed to draw any
meaningful conclusions.  The discrepant points with the unphysically
high [NII] temperatures all have very large uncertainties, and are
probably the result of the susceptibility to anomalously large 
$T_e$ values for marginally detected lines that was alluded to earlier.

The most surprising comparison is between [\oii] and [\oiii] temperatures,
as shown in the bottom panel of Fig.\ 1.  Although there may be a hint of
a correlation in the extremes of the temperature range (e.g., $T_e < 9000$ K
vs $T_e > 13000$ K), for most objects the two temperatures are uncorrelated.
In order to assure ourselves that this result was not an artifact of 
an error in our data (e.g., poorly determined reddening corrections),
we compiled data from other sources (Peimbert, Torres-Peimbert,
\& Rayo 1978; Hawley 1978; Talent \& Dufour 1979; Vermeij \& van der
Hulst 2002) and these data show a similar inconsistency in temperatures.

The source of this disagreement remains unresolved, but we can
speculate on possible sources. First, the $^2 P^o$ level that gives
rise to the [\oii]$\lambda\lambda$7319,7330 complex (actually a pair
of doublets) can be populated by direct dielectronic recombination
into that level, as well by radiative cascade (Rubin 1986). This
mechanism also affects the [\nii]$\lambda$5755 line and the
[\oii]$\lambda$3727 doublet as well, although to a lesser degree. The
net effect is that the electron temperature derived from the [\oii]
line ratio can be overestimated if the recombination contribution is
not accounted for. The magnitude of the recombination contribution
increases with both temperature and the O$^{+2}$ abundance, and so is
a greater problem for hotter nebulae with large O$^{+2}$
fractions. Liu et al.~(2000) provide correction formulae for the
recombination contributions to the [\oii] and [\nii] line intensities:

\begin{equation}
{{I_R(7319+7320+7331+7332)}\over{I({\rm H\beta})}}
= 9.36\,(T/10^4)^{0.44} \times {{\rm{O}}^{+2}\over{\rm{H}}^+},
\end{equation}
where $I_R/I({\rm H\beta})$ is the intensity of the recombination contribution to
the [\oii] line relative to H$\beta$. 

We have estimated the likely contribution of recombination to the
[\oii]$\lambda\lambda$7319,7330 feature, using the [\oiii] electron
temperatures from Table~3 and O$^{+2}$/H$^+$ ratios computed in
Table~4. We find recombination typically contributes $<$ 5\% to the
[\oii] line flux, which corresponds to a temperature error of only
$\sim$2--3\%, or less than 400~K in our worst case. Hence it cannot be
a significant contributor to the discrepancy observed here. Note that
recombination into [\oii] would be expected to scatter the derived
[\oii] temperatures to systematically higher values.

It should also be noted that the [\oii] $\lambda$7325/$\lambda3727$
ratio is much more severely affected by collisional de-excitation than
other $T_e$-sensitive ratios. A simple five-level atom analysis shows
that the derived T[\oii] can differ by as much as 1,000-2,000~K when
changing $n_e$ from 10 to 200 cm$^{-3}$.
Although our [\sii]$\lambda$6717/$\lambda$6731 line ratios typically
are consistent with $n_e$ being in the low density limit, in practice
the uncertainties in the [\sii] ratios are such that the 3$\sigma$
upper limits are of the order 200-300 cm$^{-3}$, thus permitting a
fairly considerable range in electron density.

Other physical origins for discrepant [\oii] temperatures might include
radiative transfer effects, such as part of the [\oii] emission arising
in shadowed regions preferentially ionized by the diffuse radiation
field, or a significant contribution of the [\oii] from shocked 
regions.  Shocks may be a significant contributor when one considers
the supersonic turbulence that typifies giant \hii\ regions (e.g.,
Chu \& Kennicutt 1994; Melnick, Tenorio-Tagle, \& Terlevich 1995
and references therein)\footnote{We thank the referee, Michael Dopita, 
for suggesting these mechanisms.}.

Observational uncertainties undoubtedly contribute to the scatter in
T[\oii] as well. The large wavelength difference between the [\oii]
features (3727 -- 7325 \AA) makes their ratio more sensitive to uncertainties 
in interstellar reddening. This is compounded by the fact that the two
features can not be observed in a single spectroscopic setting, because
of second-order contamination. The $\lambda$7325 feature is located in
a region of relatively strong OH airglow emission, and is in the middle 
of a telluric water vapor band as well. All these effects probably 
raise the uncertainty in the [\oii] line ratio considerably over the
formal uncertainty based on photon statistical fluctuations. It is
difficult to estimate the contributions to the uncertainties from 
some of these effects, and it is likely that the formal uncertainties 
for T[\oii] in Table~2 underestimate the true errors. 

Until the origin of the discrepancy of [\oii] temperatures with other
ions is understood, abundances derived solely on the basis of T[\oii]
measurements should be treated with skepticism. 

\subsection{Ionic and Total Elemental Abundances}

The very good correlation between T[\siii] and T[\oiii] in Figure~1,
and its consistency with the predicted relation from ionization
models, led us to calculate ionic abundances using a three-zone model
for the electron temperature structure of an \hii\ region (Garnett
1992). In this model the [\oiii] temperature is used to characterize
the highest-ionization zone (O$^{+2}$, Ne$^{+2}$), [\siii]
temperatures are used to characterize the moderate-ionization zone
(S$^{+2}$, Ar$^{+2}$), and the [\nii] and [\oii] temperatures to
characterize the low-ionization zone. However, in view of the problems
apparent in the [\oii] temperatures, we needed to adopt an alternative
procedure for setting the temperature in the low-ionization region.

We have used the scaling relations between [\oii], [\siii], and
[\oiii] temperatures from Garnett (1992) to reduce the uncertainties
in the individual $T_e$ measurements, and to estimate one or more of
the zone temperatures above when a direct $T_e$ measurement was not
available (Garnett et al.\ 1997).  Instead of applying the measured
T[\oiii] directly, we used equation [1] to derive a second estimate of
T[\oiii] and averaged the two values to set the 
temperature in the high-ionization zone.  The same process was applied
in reverse to set $T_e$ in the moderate-ionization zone.  Although
this procedure introduces some dependence between the temperatures in
the high- and moderate-ionization zones, it is justified by the tight
correlation in the top panel of Fig.~1, and it reduces the random
error in the derived abundances without introducing any systematic
errors.  We used equations [1] and [2] to estimate the temperature in
the low-ionization zone, unless a direct $T_e$ measurement was
available from [\nii].  Table~3 summarizes these adopted electron
temperatures.

Of the 22 \hii\ regions in Table 1 with auroral line measurements,
19 objects have sufficient quality temperature measurements to provide
reliable abundances.  Two regions (H237, H875) were excluded because
only [\oii] temperatures were measured, and one region (H140) was 
dropped because the remaining [\siii] or [\oiii] temperatures
were too poorly measured to provide accurate abundances measurements
on their own.  The addition of H681 from Garnett \& Kennicutt (1994)
leaves us with robust $T_e$-based abundances for 20 \hii\ regions.
Table 4 lists the ionic abundances for O$^+$, O$^{+2}$, S$^+$,
S$^{+2}$, N$^+$, Ne$^{+2}$, and Ar$^{+2}$ (all with respect to hydrogen),
computed using the adopted electron temperatures in Table 3. 
Table 5 lists the total elemental abundances for O, N, S, and Ne.

The current data set is unique in that it provides a large number of 
$T_e$-based abundances for a spiral galaxy, over large ranges in 
metal abundance and nebular ionization.  This enables us explore 
the behavior of the ionization (and the reliability of the commonly
applied ionization corrections) for several heavy elements.  We 
discuss helium separately in \S~4.4.

For oxygen, we know from the weakness of \heii~$\lambda$4686 that 
a negligible fraction of the gas is in O$^{+3}$, so the oxygen abundance 
is simply the sum (O$^+$ + O$^{+2}$). For
nitrogen we made the usual assumption that N/O scales with N$^+$/O$^+$, 
although this has not yet been demonstrated to be generally applicable
(Garnett 1990). 

Sulfur requires a correction for unobserved S$^{+3}$ in high-ionization
regions. This is illustrated in Figure~2, where we show the ratio 
(S$^+$ + S$^{+2}$)/(O$^+$ + O$^{+2}$) vs the O$^+$ fraction, for
\hii\ regions in M101 and other objects from
Garnett (1989) and Garnett et al.\ (1997).
The sharp drop in S$^+$ + S$^{+2}$ relative to O in highly-ionized
regions (i.e., low O$^+$/O)
is symptomatic of an increasing contribution from unobserved
S$^{+3}$. To account for the S$^{+3}$ contribution we used the 
correction formula of Stasi\'nska (1978) and French (1981): 

\begin{equation}
{{S^+ + S^{+2}}\over S} = 
\biggl[1 - \biggl(1- {O^+ \over O}\biggr)^\alpha\biggr]^{1/\alpha},
\end{equation}
using  $\alpha$ = 2.5, which is consistent with the sulfur ionization
trends derived from photoionization models by Garnett (1989).  This
relation is plotted as the solid curve in Figure~2, for an assumed (constant)
intrinsic log\,(S/O) = $-$1.60.  While this relation appears to adequately
follow the data for O$^+$ fractions greater than 0.2 (including 
most or all of the M101 \hii\ regions in this paper), the objects
with higher ionization appear to fall below the curve. This could
indicate that the models underpredict the S$^{+3}$ fraction in 
high-ionization nebulae.  This remains to be explored through further
modeling and observations of [S~IV]$\lambda$10.5\,$\mu$m. 
Observations of many Galactic and LMC \hii\ regions with the Infrared Space 
Observatory (ISO) indeed show strong [S\,IV] emission 
(e.g., Peeters et al.\ 2002a, b; Mart\'in-Hernandez et al.\ 2002).

Argon presents problems similar to those for sulfur, because three
ionization states can be present in \hii\ regions.  Ar$^+$ is present
in low-ionization nebulae, while Ar$^{+3}$ can appear in \hii\ regions
of the highest degree of ionization.  The ionization potentials of
sulfur and oxygen bracket the respective levels in argon, so we have tested
whether the observed ionic ratios in O or S can be used to derive 
a useful ionization correction factor (ICF) for argon.  Figure~3 (top panel)
shows that Ar$^{+2}$/O$^{+2}$ is a poor predictor of Ar/O, as this ion ratio 
is strongly correlated with O$^+$/O over the entire range sampled
by our \hii\ regions.  On the other hand a plot of 
Ar$^{+2}$/S$^{+2}$ vs O$^+$ fraction (bottom panel of Figure~3) 
shows very little correlation; only the lowest-ionization region (H336)
shows any significant deviation from a constant Ar$^{+2}$/S$^{+2}$. 
Our measurements of 15 \hii\ regions in M101 (excluding H336) and 8 
\hii\ regions in NGC~2403 (Garnett et al.~1997) yield a mean 
log\,(Ar$^{+2}$/S$^{+2}$) = --0.60$\pm$0.05. Thus for the purposes of 
this study we will assume that Ar/S = Ar$^{+2}$/S$^{+2}$. 

Neon is observed mostly in high-ionization, low-abundance \hii\ regions
via the [\neiii] $\lambda$3869 line, and in those cases it is 
usually assumed that Ne$^{+2}$/O$^{+2}$ = Ne/O.  This line  
is rarely observed in metal-rich \hii\ regions, however, so the ionization
corrections are less well constrained. A further difficulty is that models
for the ionization of neon are sensitive to the input stellar atmosphere
fluxes, which are themselves sensitive to the treatment of opacity and
stellar winds. For this reason ionization models have had difficulty
reproducing [Ne~III]. 

The determination of the form of an ICF for Ne requires infrared
measurements of [\neii] and [\neiii] lines along with optical measurements
of [\oii] and [\oiii]. Few such measurements exist over a large range
of O$^+$/O. We have examined recent results from {\it ISO} spectroscopy of
\hii\ regions in M33 by Willner \& Nelson-Patel (2002) and in the
Magellanic Clouds by Vermeij \& van der Hulst (2002). Figure~4 shows
Ne$^{+2}$/Ne vs. O$^+$/O for the regions in these studies having both
optical and IR measurements.  The data clearly indicate that
Ne$^{+2}$/Ne declines strongly as the O$^+$ fraction increases. On the
other hand, the observational scatter is significant, and there are
too few measurements for low-ionization regions (O$^+$/O $>$ 0.5) 
to reliably derive a
functional form for the Ne ICF at this time.  Figure~5 shows the
Ne$^{+2}$/O$^{+2}$ ratios for our M101 \hii\ regions.  The data are
consistent with a constant ratio over a wide range of ionization,
but the scatter increases considerably as O$^+$/O increases. This is
partly due to observational scatter, but probably also reflects
greater sensitivity to the local radiation field as Ne$^{+2}$ and
O$^{+2}$ become minor constituents.  The data are consistent with a
constant Ne/O abundance ratio in M101, but do not establish this
unambiguously.  In view of the uncertainty in the Ne ICFs we shall not
discuss the neon abundances further in this paper.

\section{The Abundance Gradient in M101}

These new abundance measurements provide largest and most complete 
radial coverage of any galaxy measured to date.  We first discuss
the radial behavior of the overall heavy element abundance gradient
(via O/H), and then examine the other heavy element and helium
abundance properties.

\subsection{Oxygen Abundances}

Figure~6 shows the radial dependence of the oxygen abundances for
the 20 \hii\ regions in our sample with reliable $T_e$ measurements.  
The error bars show the uncertainty in
abundance as propagated from errors in the line measurements; these
are dominated by the auroral line measurements.  
For most of the \hii\ regions in the sample, the choice of ionic temperature
for the low-ionization zone has a small effect on the derived abundance
($\le$0.05 dex).  This is a reflection of the consistency of the [\siii]
and [\oiii] temperatures discussed earlier, combined with the milder
dependence of abundance on $T_e$ for intermediate-abundance \hii\ regions.
However the choice of temperature is significant at the extremes of 
the abundance distribution, especially for the three innermost \hii\ regions.

In terms of the revised solar value of 12 + log\,(O/H)$_\odot$ = 8.7,
the gas-phase oxygen abundances in M101 range from 1.2 $\pm$
0.2\,(O/H)$_\odot$ in the center to 0.07 $\pm$ 0.01\,(O/H)$_\odot$ at R
= 41 kpc.  The abundance in the outermost disk is comparable to that
of all but the most extreme metal-poor dwarf irregular galaxies.  For
comparison I Zw 18, the most metal-poor \hii\ galaxy known, has an
abundance 12 + log\,(O/H)~=~7.2 or 0.03\,(O/H)$_\odot$ on the new solar
scale.

Numerous other authors have measured $T_e$-based abundances for small
samples of \hii\ regions in M101, as compiled in Table~6.  Figure~7
shows that our results are generally in good agreement with these
previous measurements.  The rms difference between our O/H values and
those from the literature is 0.11 dex, or 0.09 dex if we exclude H336,
the most discrepant point, and we see no large-scale systematic
differences.  As one might expect the differences are larger for the
more metal-rich \hii\ regions, which have lower and more difficult to
measure temperatures.  The abundances for these regions are very
sensitive to the adopted $T_e$; for example in the innermost region
H336 a difference in $T_e$ of only 600~K in T[\oii] accounts for about
0.2 dex of the difference between our abundance and that of Kinkel \&
Rosa (1994). The rest comes from their use of an ionization model to
derive O/H; if we use their measured $T_e$ values, we derive 12 +
log\,(O/H) = 8.79.

The observed abundances are well fitted by an exponential distribution,
as shown by the solid line in Figure~6.  This line shows a least squares
fit to the points, yielding:

\begin{equation}
\rm 12 + \log\,(O/H) = 8.76\,(\pm 0.06) - 0.90\,(\pm 0.08)\,(R/R_0). 
\end{equation}

An independent $T_e$-based abundance measurement to H336 (Searle 5)
is available from Kinkel \& Rosa (1994).  Those authors derived 
12 + log\,(O/H) = 8.88 (no uncertainty quoted), which is significantly
higher than our value of 8.55$\pm$0.16.  Their measurement is plotted
as an open circle in Figure~6, and the dashed line shows the resulting
fitted exponential abundance distribution if we use their measurement
instead of ours:

\begin{equation}
\rm 12 + \log\,(O/H) = 8.85\,(\pm 0.07) - 1.02\,(\pm 0.10)\,(R/R_0).  
\end{equation}

The difference propagates
to an uncertainty of $\pm$0.1 dex in the extrapolated central abundance
At the adopted distance of M101 (7.5 Mpc), $R_0$ = 32.4 kpc, so
the corresponding gradients from equations [5] and [6] are 
$-$0.029 and $-$0.032 dex kpc$^{-1}$ respectively. 
of M101 and $\pm$10\% in the slope of the gradient.
For a B-band disk scale length of 5.4 kpc (Okamura, Kanazawa, \& Kodaira 1976), 
the corresponding gradient is $-$0.16 $\pm$ 0.01 dex per unit scale length.

The fact that the gradient in O/H remains closely exponential over six
disk scale lengths is remarkable. There have been claims for a change
in slope of the gradient in M101 at 0.3--0.5~$R_0$, based on
strong-line estimates of O/H, and it has been argued that such a
change could be related to the change in slope of the rotation curve
(Scowen et al.~1992; Zaritsky 1992).  More recently Henry \& Howard (1995)
and Pilyugin (2003) have argued that an exponential
gradient was the best fit to the published data.  Our results confirm
that the purported changes in gradient slope are spurious and probably a result
of systematic errors in the strong-line abundance estimators
(\S~5). Claims of changes in the slope of abundance gradients based
only on strong-line estimates should be regarded as uncertain.

Exponential heavy element abundance gradients have important 
consequences for chemical evolution models.
A generic feature of chemical evolution models is that the composition
gradients flatten with time, as a result of the breakdown of the
instantaneous recycling approximation; as old, metal-poor stars die,
they eject relatively metal-poor gas into the ISM at late times, and
thus inhibit the enrichment of the ISM. This process proceeds more
quickly in the dense inner regions of spirals, so that the models
predict a flatter gradient in the inner disk than in the outer (e.g.,
Prantzos \& Boissier 2000; Chiappini et al.~2002). However, such
flattening of the gradients is rarely observed. This implies that some
input assumption of the models is inadequate. Two possible
explanations are that: (1) Disks are younger than
assumed in the published chemical evolution models. In the models, the abundance
gradients steepen with time early on, and then flatten from the
inside out as the galaxy ages.  In young disks the composition
gradients are close to exponential.  Thus if disks are younger than often
assumed in the models (13-15 Gyr), then more uniform gradients would
be expected. (2) Slow radial gas flows may redistribute gas and metals
so as to maintain an exponential composition gradient.  Lin \& Pringle
(1987) showed that viscous evolution of matter and angular momentum in
disks can lead to exponential density distributions, while Clarke
(1989) and Sommer-Larsen \& Yoshii (1989) demonstrated that this can
give rise to exponential composition gradients as well.  It is as yet
unclear how such radial gas flows can be maintained, or even what the
magnitude of radial flows could be.  Numerical simulations 
should help to address these questions (e.g., Churches, Nelson, \& 
Edmunds 2001).  If steep exponential abundance gradients are a 
common feature in disks this will need to be taken into account in galaxy
evolution models.

Despite the well-defined exponential distribution, the points in
Figure~6 show considerable scatter. Inspection of the data on
a point-by-point basis shows that most or all of this scatter is
due to uncertainties in the individual abundance measurements.  
For example, the scattering of points near $R/R_0$ = 0.42 and 0.53
are mainly different emission regions in the NGC~5462 and NGC~5447
complexes, respectively, which presumably have the same physical abundances.  
The rms residual between the data points and the fit is $\pm$0.09 dex,
which is only marginally larger than the rms uncertainty of the 
individual measurements ($\pm$0.07 dex), and only one object (H128) deviates 
by more than 3$\sigma$ from the fitted gradient.  As a result our
data set is neither large enough nor accurate enough to test for 
local variations in abundances, such as the azimuthal asymmetry in 
nebular excitation and empirical abundance that was reported in Paper~I.

\subsection{N/O Abundance Ratios}

Nitrogen shows a strong radial abundance gradient as well, and 
the behavior of the the differential N/O abundance ratio can
place contraints on its nucleosynthetic origin.

Figure~8 shows the behavior of N/O with galactocentric radius 
(top panel) and as a function of oxygen abundance (bottom panel).
There is a strong gradient in N/O across M101, as suspected from 
previous studies, and a significant correlation of N/O with overall O/H
abundance. 
The two outermost \hii\ regions in M101 have N/O values similar to
those seen in dwarf irregular galaxies (e.g., Garnett 1990). 
The data in Figure~8 suggest a possible flattening of the N/O gradient 
in the outer parts of M101, consistent with the roughly constant value
observed in dwarf galaxies, but we do not have enough 
data points to draw strong conclusions.

One region, NGC~5471-C, stands out as having a significantly higher
N/O than expected. This is not an artifact of the data, as we saw
relatively strong [\nii] in our spectrum of this position from 
Paper~I, and a similar enhancement was seen in observations of 
the region by Skillman (1985). 
NGC~5471 is known to contain a luminous supernova remnant at the position
of NGC~5471-B (Skillman 1985; Chu \& Kennicutt 1994; Chen et al.~2002), 
which can enhance the [\nii] and [\sii] lines.  High resolution
\halpha\ and [\sii] images of NGC~5471 from {\it HST} do not show any
anomalous structure in NGC~5471-C (Chen et al.\ 2002), 
but echelle observations by
Chu \& Kennicutt (1986) do reveal a region of high-velocity gas
located within 0\farcs3 of the object.  These fast-expanding regions
often are associated with supernova remnants (e.g., Chu \& Kennicutt
1994).  Another possibility is a selective enhancement of nitrogen by
the presence of Wolf-Rayet stars (Kobulnicky et al.~1997). 

There is some debate over the existence of an N/O gradient within the Milky
Way disk, with optical measurements giving relatively shallow N/O gradients
while infrared studies show steeper gradients (cf.~Shaver et al.~1983 vs 
Lester et al.~1987). It is not clear whether this discrepancy is 
due to ionization effects or observational errors (Garnett 1990). 
But the N/O gradient in M101 is unmistakable (Fig.~8), even without
infrared measurements.  

It is also instructive to compare the behavior of the N/O abundances
in M101 with expectations from chemical evolution models. In the 
simple instantaneous chemical evolution model (Pagel 1997),
`primary' elements (i.e., those that are the ultimate product of the 
original H and He in a star at birth), are expected to vary in lockstep.
A `secondary' element (i.e., one produced from a heavy element seed
present in the star at birth), on the other hand, is expected to grow
in proportion to the abundance of primary elements. The solid line in 
the bottom panel of Figure~8 shows a simple model which includes a 
component with constant log\,(N/O) = --1.5, representing primary 
nitrogen, and a component for which log\,(N/O) = log\,(O/H) + 2.2, 
representing secondary nitrogen production. This simple representation 
provides an adequate description of the M101 data. 

For comparison the dashed line in Fig.~8 shows a numerical 
one-zone model for the
evolution of N/O vs. O/H from Henry, Edmunds, \& K\"oppen (2000;
hereafter HEK00).  The line corresponds to their ``best'' Model B,
which best reproduces the trend of N/O vs O/H in the data set they
compiled for the study.  Surprisingly, this model fails to reproduce
our M101 results; the predicted N/O is too low at intermediate
abundance (log\,O/H = 8.2 to 8.6), and rises too steeply at high
abundance.  Comparison with their other Models A and C (with lower and
higher star formation efficiencies, respectively) does not alter this
conclusion. We suspect that this discrepancy is mainly due to the
oxygen abundance scale adopted by HEK00.  Their data for spirals
consisted largely of \hii\ regions with abundances derived from
strong-line calibrations, which yield systematically higher O/H than
$T_e$-based abundance measurements (\S5). In contrast, N/O is only
modestly affected by the different abundance determination methods,
because of its much weaker dependence on electron temperature. 
The result is that we
observe N/O to increase more steeply at intermediate O/H than the
HEK00 models.  This example illustrates the critical importance of
robust abundance measurements of extragalactic \hii\ regions, and of
establishing the normalization of the nebular abundance scale, for
modeling and understanding the chemical evolution of galaxies.

On the basis of their models, HEK00 suggested that the secondary component 
of nitrogen in spiral disks arises mainly from processing of carbon,
that is, mainly from CN cycling in intermediate mass stars. On the
other hand, our M101 data suggest that nitrogen enhancement (relative
to oxygen) is inititated at much lower metal abundances than in
the HEK00 models. The fact that we can reproduce the trend
in N/O with a component that grows proportionately with O/H 
suggests that secondary nitrogen is produced mainly from oxygen, and implies
that the main source is full CNO cycling in massive stars. A comparison of
the variation of nitrogen with carbon and oxygen abundances would
be very informative in this regard, but this awaits
the acquistion of much more data on carbon abundances in galaxies.

\subsection{S/O and Ar/O Abundance Ratios}

Elementary nucleosynthesis theory predicts a common primary origin for oxygen,
sulfur, and argon in the cores of massive stars, so one would expect
much less systematic variation in the S/O and Ar/O abundance ratios.
Figure~9 shows the behavior of S/O and Ar/O in M101 (where Ar/O is
derived from Ar/S and S/O, as discussed in \S~4.2).
Excluding the lower limit for the outermost region (SDH~323), we see
no dependence of S/O on absolute abundance.  The two most metal-rich
regions in our sample do appear to have systematically smaller Ar/O
than the bulk of the sample, and it is not clear whether the low
values might be the result of observational errors or uncertainties in the
ionization correction. The dashed lines in Figure~9 show the average
S/O and Ar/O ratios for \hii\ regions in dwarf irregular galaxies (as
compiled by Izotov \& Thuan 1999). Our results are consistent with the
extrapolation of these trends, although our average S/O appears to be
slightly smaller (and our average Ar/O slightly larger) than those
reported by Izotov \& Thuan.

Observations of M51 by D\'\i az et al.~(1991) suggest evidence for a
lower S/O ratio in very metal-rich \hii\ regions. We do not see any of
a similar decline in S/O in the inner disk of M101, but our
measurements do not extend to as high metallicities as in M51.  On the
other hand, we do see hints of a lower Ar/O at high metallicity in
M101.  It should be noted that if the electron temperatures are higher
in the M51 regions than those predicted by the photoionization models
(\S5), then D\'\i az et al.~(1991) would have derived higher S/O
ratios.  A decrease in S/O and Ar/O in metal-rich environments would
be of considerable interest, as it would indicate some dependence of
stellar nucleosynthesis or the stellar initial mass function with
metallicity.

\subsection{Helium Abundances}

The large abundance range in our data set allows us to 
test for a systematic change in the helium abundance with
metallicity.  Although our data were not taken for this purpose, we
have high $S/N$ measurements of several \hei\ lines for many of the
\hii\ regions, as summarized in Table~7.  Listed are the \hei\ line
strengths (relative to H$\beta$), along with the corresponding
uncertainties and equivalent widths (EWs). The \hei\ lines are
influenced by underlying stellar absorption, and this will introduce
spurious gradients in the derived He abundances if corrections for the
absorption are not applied.  Olofsson (1995) computed equivalent
widths of stellar \hi\ and \hei\ lines for OB star associations;
however, these calculations do not include \hei\ from B stars. This is
unfortunate since the \hei\ EWs in stars are a maximum at spectral
type B0. Gonz\'alez-Delgado, Leitherer, \& Heckman (1999) model
starbursts consistently over a wide range of ages, but provide \hei\
EWs only for lines blueward of 5000 \AA. Self-consistent calculations
of \hi\ and \hei\ EWs for OB associations for the stronger red \hei\
lines are needed to estimate corrections for He abundances in \hii\
regions. This is especially important at low metallicities for the
primordial He problem (e.g., Olive, Skillman, \& Steigman 1997; Izotov
\& Thuan 1999).

We derived approximate corrections for \hei\ absorption based on
direct measurements of \hei\ EWs for stars by Conti \& Altschuler
(1971), and Conti (1973, 1974), which provide a comprehensive
compilation of He EWs as a function of spectral type for O
stars. Guided by the results of Gonz\'alez-Delgado et al.~(1999) for
the \hei\ $\lambda$4026 and $\lambda$4471 lines in 2-3 Myr old
starbursts, we assumed a luminosity-weighted average spectral type of
O7 for our M101 regions and obtain estimated EWs of 0.5 \AA\ for
$\lambda$4026, 0.5 \AA\ for $\lambda$4471, 0.8 \AA\ for $\lambda$5876,
and 0.3 \AA\ for $\lambda$6678.  In computing He abundances we used
the \hei\ emissivities of Benjamin et al.~(1999).  Correcting our
\hei\ emission line strengths for these absorption EWs yields much
better agreement between the He$^+$ abundances computed from different
lines. The He$^+$ abundances computed from each measured \hei\ line
are listed in Table~8, along with the weighted average for all
measured \hei\ lines in each \hii\ region.  Note that the He$^+$
abundances for the outermost region SDH~23 differ considerably for the
4471 and 5876 \AA\ lines; the 5876 \AA\ line is affected by a cosmic
ray, and so we adopt the value derived from $\lambda$4471 as the
He$^+$ abundance for this object.

The final correction in our derivation of He/H is accounting
for neutral He, which is not directly observable in the ionized region.
As shown in Figure~10 (open circles), the derived He$^+$/H$^+$
abundance ratio declines in the metal-rich inner disks of M101 
and other spiral galaxies (also see BKG). 
This is probably due to the presence of 
cooler ionizing stars and low ionization in these \hii\ regions (BKG).
We estimated ionization correction factors for He using the ionization
models of Stasi\'nska (1990).  The He$^+$ fraction is roughly 
correlated with O$^+$/O, with the He$^+$ fraction close to unity
for O$^+$/O $<$ 0.2, and He almost entirely neutral for O$^+$/O
$>$ 0.9. At intermediate values, however, the He$^+$ fraction can
vary considerably at fixed O$^+$/O due to variations in ionization
parameter, making the He ICF rather degenerate. This degeneracy can
be broken if the ionization parameter can be constrained, for instance
using the S$^+$/S$^{+2}$ ratio. We therefore use our measured values
of S$^+$/S$^{+2}$ and O$^+$/O to estimate more accurately the ICFs
for He.

The resulting He abundances are plotted as a function of radius in
Figure~10 (solid points), and as a function of O/H in Figure~11.  Although the
measurement uncertainties are substantial, the plots show the presence
of a significant radial gradient in He/H, after the neutral He
corrections are applied.  Only the outermost region (SDH~323) shows
much deviation from this gradient, but its measurement is very
uncertain due to the lack of a [\siii]$\lambda\lambda$9069,9532
measurement for this object.
The trend suggests we have likely overestimated the ICF for SDH~323,
but this should be confirmed by observations of [\siii].

The solid and dashed curves in Figure~11 show fits by  
Olive et al.~(1997) to He abundance measurements in two samples of 
metal-poor dwarf galaxies (their sets B and C, respectively;  
see their Table~1 for parameters of the fits). The M101 He abundances
appear to extrapolate the dwarf galaxy trends very well. We can not
yet clearly differentiate between the two fits, although our
abundances appear to be more consistent with the shallower dependence
on O/H. A larger set of precise He line measurements for metal-rich
regions would provide better limits on the slope of the He/O relation,
which is an important constraint on stellar nucleosynthesis and the
effects of stellar mass loss on yields (Maeder 1992), as well as on
the primordial He abundance.

\section{Comparison with Empirical Abundance Scales}

As discussed in \S1, abundance estimates based on strong-line
``empirical" methods have largely supplanted direct, $T_e$-based
determinations for large-scale abundance surveys and cosmological
lookback studies.  With the large dynamic range in O/H in our data
set, it is instructive to compare our direct abundances with those
derived for the same \hii\ regions using the most popular strong-line
calibrations.  In order to enlarge the comparison we have also
included recent high-quality direct abundance measurements of \hii\
regions in other galaxies from the literature (Garnett et al.~1997,
van Zee et al.~1998; Deharveng et al.~2000; D\'\i az \&
P\'erez-Montero 2000; Castellanos et al.~2002).

Figure~12 shows the well-known plot of oxygen abundance against the
empirical abundance index $R_{23}$ (Pagel et al.~1979), where:

\begin{equation}
R_{23} \equiv {{(I([\oii]\lambda\lambda3726,3729) +
 I([\oiii]\lambda\lambda4959,5007))} \over {I(H\beta)}}.
\end{equation}

\hii\ regions from this paper are plotted with solid points, while objects
from the other papers cited above are indicated with open squares.  The
superimposed lines show examples of widely-used calibrations for 
$R_{23}$, as taken from the references indicated in each panel.  These
particular examples were selected to illustrate the range of $R_{23}$ 
calibrations in the literature.  
The $R_{23}$ index shows the well-known double-valued behavior, with
oxygen line strength increasing monotonically with abundance at low
metallicities, and decreasing at high metallicities, the latter reflecting the
dominance of oxygen cooling over abundance in metal-rich regions.
Since most of the \hii\ regions observed in spiral galaxies have
metallicities that place them on the ``upper" branch of this diagram,
we will restrict the discussion in this paper to that part of the
strong-line calibration, roughly corresponding to $12 + \log~(O/H) \ge 8.0$.

Although the measured \hii\ region abundances trace a locus which is 
roughly consistent in shape with most of the empirical calibrations, there
is a pronounced offset in abundance in most cases, as has been 
pointed out previously (see Stasi\'nska 2002 and references therein).
In all cases the empirical calibrations yield oxygen abundances that
are systematically higher than the $T_e$-based abundances, by amounts
ranging from 0.1 to 0.5 dex, depending on the calibration and the
excitation range considered.  
The calibrations of Pilyugin (2000, 2001a) shown in the lower right
panel of Figure~12 show a smaller offset, mainly because those calibrations
are partly based on $T_e$-based abundances.

Similar discrepancies are seen in most of the other ``empirical"
abundance indices, as is illustrated in Figure~13.  Here we directly
compare the $T_e$-based abundances with those derived from several
different empirical indices, all 
calibrated with a common grid of photoionization models 
(Kewley \& Dopita 2002), so the different indices can be evaluated
on the same basis.  The symbols are the same as in Figure~12.
As noted earlier, there is a systematic shift toward larger abundances
in the empirical calibrations, ranging from 0.2 to 0.5 dex on average.

The discrepancies shown in Figures 12 and 13 can be traced to two
main origins, an insufficient number of calibrating \hii\ regions
with accurate $T_e$-based abundances in the earliest calibrations,
and a systematic offset between the nebular electron 
temperatures in the calibrating photoionization models and the observed
forbidden-line temperatures, for a given strong-line spectrum
(e.g., Stasi\'nska 2000).  Because of the exponential temperature dependence
on the strengths of these collisionally-excited forbidden lines  
($\propto \exp{-\chi/{kT}}$), an offset in $T_e$ will have the strongest effect
on the shorter wavelength features at a given abundance.  The better
consistency of $S_{23}$ in Figure~13 can be attributed in part to the
relatively low excitation of the [\siii]$\lambda\lambda$9069,9532 
lines that dominate this index; note that at higher abundances,
where the characteristic electron temperatures are very low, even
this index deviates significantly from the direct abundance determinations.

Figure~14 shows a comparison of direct abundances with empirical
values derived using the ``$P$ method" of Pilyugin (2001a).  At
intermediate abundances (12 + log\,(O/H) $\simeq$ 8.5) the two abundance
scales are in approximate agreement, in part because the method is
calibrated with direct abundance measurements in this region.  However
the $P$ method results break down for 12 + log\,(O/H) $\le$ 8.4.
As can be seen in Figure~12, these regions fall into the
abundance range where most of the empirical indicators become
multiple-valued or very sensitive to nebular
ionization parameter.  Most authors are aware of these limitations and
restrict the application of the various empirical tracers to the
abundance range for which they are valid (see Pilyugin 2001a; Kewley
\& Dopita 2002 for detailed discussions).  However the comparison in
Figure~14 illustrates a pitfall in this approach.  The $P$ method as
calibrated is supposedly only valid for 12 + log\,(O/H) $\ge$ 8.2
(Pilyugin 2001a), but in the absence of electron temperature
measurements when does one know for sure that an \hii\ region falls
within the applicable abundance range?  If the empirical abundances
overestimate the actual abundances by a significant factor, they may
be applied in regions for which the calibration is invalid.  

\section{DISCUSSION}

The results in the previous section may
have significant consequences for the nebular abundance
scale as a whole.  If the forbidden-line abundances are correct, it
implies that most studies of galactic abundances (locally and at high redshift)
based on ``empirical" nebular calibrations have over-estimated the true 
absolute oxygen abundances by factors of 1.5--3, for 
$12 + \log ({\rm O/H}) > 8.2$ (about 1/3 Z/Z$_\odot$).  
Our results do not apply to the low-metallicity part of the abundance
scale, so another net effect of this change would be to reduce
the total range of nebular abundances observed in galaxies.  This
may have significant effects on the measured abundance gradient
slopes in some galaxies, and on applications such as the metallicity
dependence of the Cepheid period-luminosity relation (e.g., Kennicutt
et al.\ 1998).

It is not completely clear whether the discrepancies in Figures 12 and 13 are  
the result of a systematic error in the $T_e$-based abundances,
the model-based ``empirical" abundances, or both.  
There are well-documented reasons for questioning the accuracy of 
both abundance scales (also see Peimbert 2002, Stasi\'nska 2002, 
and references therein).  

It has been argued for many years that directly measured electron
temperatures and the corresponding abundances from collisionally-excited
lines may have systematic errors due to temperature fluctuations (Peimbert
1967).  Because of the exponential temperature dependences of 
forbidden line emissivities, their fluxes will tend to be weighted toward
regions of higher temperature, and in the presence of significant 
$T_e$ fluctuations they will yield anomalously high $T_e$ measurements,
and thus low abundances. 
The importance of this effect on the nebular abundance scale depends
on the magnitude of the actual temperature fluctuations in
extragalactic \hii\ regions.  Observational evidence for significant
$T_e$ fluctuations has been contradictory. Early comparisons of
T[\oiii] in the Orion nebula with the temperature derived from the 
Balmer continuum [T$(Ba_c)$] revealed a large
discrepancy, but this has not been confirmed by subsequent 
measurements with improved detectors (Liu et al.~ 1995a).  Significant
differences between T[\oiii] and T$(Ba_c)$ are observed in planetary
nebulae (Liu et al.~2001), but it is not clear whether the two temperatures
should be the same, given temperature stratification in
ionized nebulae (Stasi\'nska 1978; Garnett 1992). Direct searches for
small-scale temperature fluctuations using high resolution {\it HST}
imaging of planetary nebulae have yet to reveal any significant
variations (Rubin et al.~2002).

Measurements of recombination lines of heavy elements (\oii, \cii, \nii)
provide an independent means of measuring nebular abundances, and they
tend to yield higher values than the corresponding forbidden lines,
by a factor that varies considerably (Liu et al.~1995b, 2001; 
Garnett \& Dinerstein 2002).  Recently Esteban et al.~(2002) 
reported measurements of the \oii\ recombination lines in the 
M101 \hii\ regions NGC~5461 (H1105) and NGC~5471.  These yielded
O$^{+2}$/H abundances that are 0.34~dex and 0.15~dex higher
than our forbidden-line abundances, respectively, and this is consistent
with the trends observed in other \hii\ regions and planetary nebulae.
However in many objects the abundance discrepancies are so
large that they can not be explained by temperature fluctuations.
The real explanation is not yet clear; Liu et al.~(2001) postulate the
presence of cold, dense, H-poor knots which produce most of the enhanced
recombination line emission, while Garnett \& Dinerstein (2002) note that
dielectronic recombination may play a significant role.  Unfortunately 
dielectronic recombination rates are rather uncertain at present (Savin 2000). 

A third concern is relevant to metal-rich \hii\ regions. At high
abundances, strong cooling by infrared fine-structure lines of [\oiii] and
[\niii] greatly depresses the electron temperature in the high-ionization
zone of \hii\ regions relative to the lower-ionization zones, leading 
to a strong temperature gradient (Stasi\'nska 1978). Under these
conditions, the measured electron temperature can be higher than the
ion-weighted mean temperature for a given species, and the optical 
forbidden lines can be weighted toward the higher temperature zones, 
leading one to underestimate the true abundances. The temperature
gradient mimics the effects of temperature fluctuations, and can
introduce differences between the measured $T_e$ and the ion-weighted
temperature of up to 2000~K (see Fig.~6b in Garnett 1992).  
The results are sensitive to ionization parameter (which
affects the amount of O$^{+2}$, and density (which affects the cooling
rate in the infrared lines through collisional quenching). 

Despite these concerns, however, there are good reasons to believe
that the forbidden lines give close to the correct abundances, at
least for the range of O/H of interest here. Recent measurements of 
infrared forbidden lines (which have weak temperature dependences)
in the spectra of planetary nebulae  
yield abundances that are in good agreement with those from optical 
forbidden lines, even where there is a large discrepancy with 
recombination line abundances (Liu et al.~2000, 2001). 
Likewise, measurements of radio recombination line temperatures
of Galactic \hii\ regions give results that are consistent with
forbidden line measurements of the same objects (Shaver et al.~1983; 
Deharveng et al.~2000).

Additional support for the reliability of the
forbidden-line abundance scale comes from independent measurements of
the local interstellar O/H from Galactic \hii\ regions and
ultraviolet absorption line measurements of the diffuse ISM.  A recent
determination of the Galactic abundance gradient by Deharveng et
al.~(2000), based on [\oiii]$\lambda$4363 measurements for 6 \hii\
regions with $R_G = 6.6 - 14.8$ kpc, yields a local oxygen abundance 12
+ log\,(O/H)~=~8.48.  This value is in excellent agreement with local
diffuse abundances of 8.50$\pm$0.02, 8.48$\pm$0.03, and 8.54$\pm$0.01
derived from {\it HST} and {\it FUSE} spectroscopy of multiple local sightlines
(Meyer, Jura, \& Cardelli 1998; Moos et al.\ 2002).  These values
are also in reasonable accord with the solar abundance (8.7), assuming
modest grain depletion factors.

If the discrepancies in Figures 12--13 are due to errors in the 
model-derived ``empirical" abundances, then what might be the cause?
The typical \hii\ region model is spherically
symmetric, with a uniform density of 100 (or 10) cm$^{-3}$ and a central
point ionizing source. In reality, the density distribution in 
many giant \hii\ regions is dominated by filamentary, shell-like
structures, often with dispersed, multiple ionizing sources within
the boundaries of the ionized gas (e.g., Kennicutt 1984).  
In moderate to high-abundance
\hii\ regions the cooling is dominated by a few ions such as O$^{++}$,
and this introduces a physical coupling between the thermal and
ionization structures, which in turn can be influenced
by the detailed distributions of gas and ionizing sources.

Unfortunately there are relatively few direct measurements of
the electron density distributions, or even the characteristic
densities in these regions.  Density measurements from the 
[\sii]$\lambda\lambda$6717,6731 line ratios generally lie at the 
low density limit, indicating 
$N_e < 200$ cm$^{-3}$.  Corroborating density measurements
from other diagnostic ratios such as [\oii]$\lambda\lambda$3726,3729 are rare. 
Accurate knowledge of the gas densities are particular important
for metal-rich regions, 
because of the dramatic effect of collisional quenching on fine-structure 
cooling and the thermal structure of the nebulae (Oey \& Kennicutt 1993).

The geometry of the ionizing sources within this complex structure
may also influence the emergent spectra.  The ionizing 
star clusters in some regions are centrally concentrated (e.g., 30 Doradus
in the LMC, NGC~5461 in M101), 
but in others (e.g., NGC~604 in M33, I Zw 18), the ionizing stars are loosely
distributed. Even in the 30 Doradus nebula, O3 stars are known to be 
located among the ionized filaments (Walborn 1991, Bosch et al.~1999). 
Since the ionization
parameter depends in part on the distance of the ionizing source from
the gas, an extended distribution of stars might be expected to affect
the resulting nebular spectrum. Dust also affects the ionized
gas, depleting Lyman-continuum photons, and by photoelectric heating
and recombination cooling of the gas. 

Finally, uncertainties in the input ionizing continua may play a
role. Although the situation is
improving here, with increasing use of non-LTE, spherical calculations
which include realistic opacities and stellar winds, models for
Lyman-continuum fluxes are still in the development stage. As shown by
BKG and Bresolin \& Kennicutt (2002), the improper
treatment of the stellar ionizing flux, especially during the
Wolf-Rayet phase, can lead to large discrepancies between observed
\hii\ spectra and those predicted by photoionization models based on
synthetic spectral energy distributions.
This situation is improving, thanks to advances in the codes describing
massive star atmospheres.  Recent stellar models 
(Martins, Schaerer, \& Hillier 2002; Herrero, Puls, \& Najarro 2002)
show that the inclusion of mass loss and more realistic stellar opacities 
has an important effect on the effective temperature scale for O stars. 
Evolutionary models based on these new stellar atmospheres by Smith,
Norris, \& Crowther (2002) appear to be in good agreement with the
ionization properties of extragalactic \hii\ regions. Further study
should determine if the directly measured and predicted \hii\ region
sequences are consistent.

In summary, we cannot be certain whether the discrepancies in 
abundance scales summarized in Figures 12--13 are due to biases
in the $T_e$-based results, or problems in the theoretical models
that are used to calibrate most of the strong-line ``empirical"
abundance scales.  High-quality far-infrared measurements of 
a sample of extragalactic HII regions, including some of the principal
fine-structure cooling lines, may help to resolve these inconsistencies.
In the meantime it is important for users 
of the various abundance calibration methods to be aware of this 
discrepancy, and until it is resolved users should beware of combining 
strong-line abundance estimates with direct abundance measurements.

An accurate extragalactic metallicity scale is fundamental for
constraining models of chemical enrichment, the chemical 
evolution of galaxies, and the cosmic baryon cycle.  In this context
it is sobering to recognize that there is a systematic uncertainty of
a factor 2--3 in the zeropoints of one or more of the metal abundance 
calibrations. A few specific
observational and theoretical investigations would go far toward
resolving this problem.  High signal/noise spectroscopy and electron
temperature measurements for other extragalactic \hii\ regions, especially 
metal-rich objects, would nail down the $T_e$-based abundance scale
across the full O/H range, and help to quantify the disagreement
between observations and ionization models across the full spectrum
of abundances, densities, and ionization properties.  Measurements
of multiple auroral lines in the same objects (e.g., Fig.~1) would
be especially valuable for reducing systematic errors in the 
$T_e$-based abundances, testing the reliability of specific nebular
thermometers, and comparing with the predictions of the nebular models.
Independent temperature and abundance measurements for \hii\ regions,
and extension of the wavelength coverage to the mid- and far-infrared
emission lines would also help.  And comparisons with young stellar
abundances will provide independent constraints on the metallicity
zeropoints (e.g., Venn et al.\ 2000; Trundle et al.\ 2002). 
On the theoretical side, explorations
of the effects of nebular geometry, structure, and ionizing spectrum
on the thermal structure of \hii\ regions would be useful, along
with targeted studies of specific transitions such as the [\oii]
thermometer (\S3.1).  Solving the abundance zeropoint problem is
important not only for the observations of metallicities of galaxies,
but also for validating our theoretical modeling and understanding
of photoionized regions in galaxies.

\acknowledgements

We are very grateful to Gary Ferland, Lisa Kewley, Manuel Peimbert
for valuable discussions about this subject.  We gratefully acknowledge
the referee, Mike Dopita, for insightful comments that improved the
paper significantly.  We also thank Dick
Henry for sending an electronic version of the model shown in Figure~8.
RCK acknowledges the support of the NSF through grant AST98-11789
and NASA through grant NAG5-8426. DRG is supported by NASA grant 
NAG5-7734.

\clearpage

\begin{figure}
\vspace{16.0cm}
\includegraphics{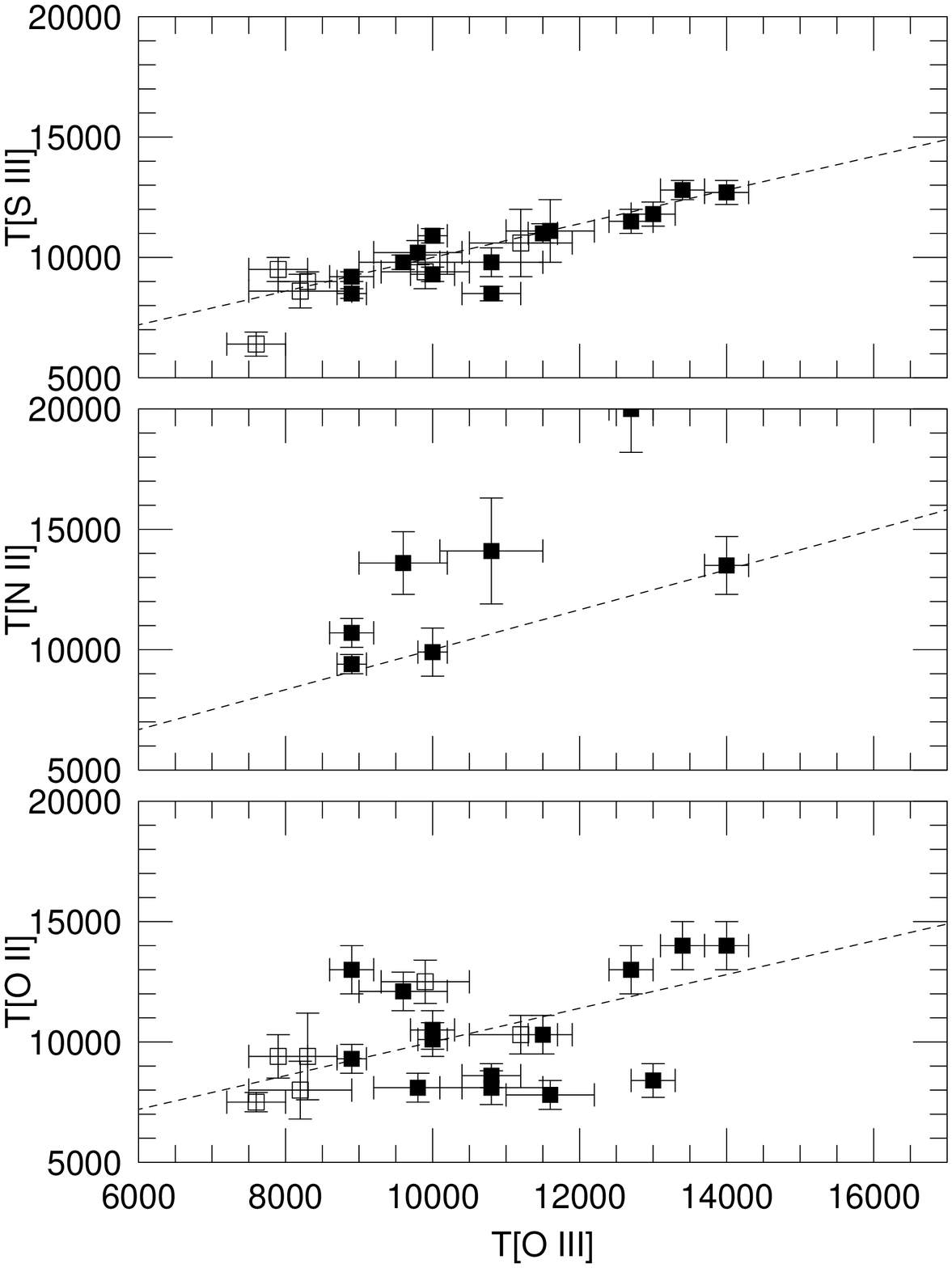}
\figcaption{The correlations between various emission-line electron temperature 
diagnostics.  
Filled symbols show our measurements for M101 \hii\ regions; open symbols
show measurements for \hii\ regions in NGC~2403 from Garnett et al.~(1997). 
The dashed line shows the predicted correlations from Garnett (1992) based
on photoionization model calculations.
}
\end{figure}

\clearpage 

\begin{figure}
\plotone{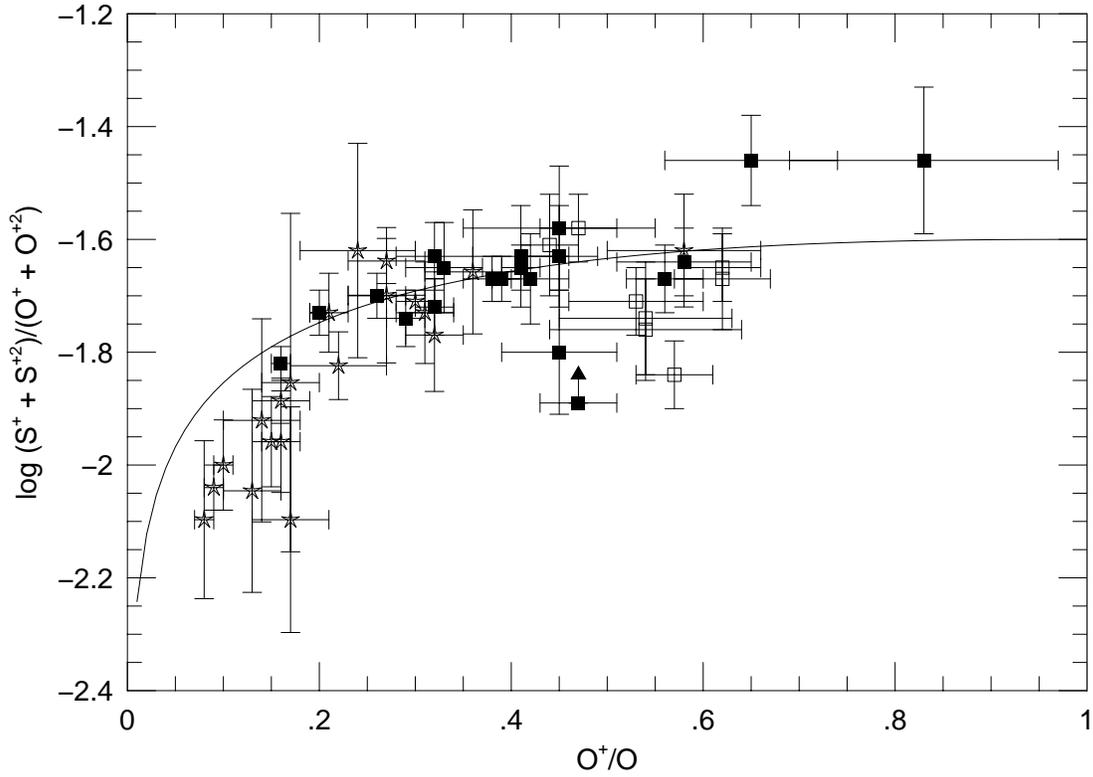}
\figcaption{The ion ratio (S$^+$ + S$^{+2}$)/(O$^+$ + O$^{+2}$) as 
a function of oxygen ionization (O$^+$/O) for
our M101 \hii\ regions ({\it filled squares}). Also plotted are data
for NGC 2403 \hii\ regions ({\it open squares}; Garnett et al.~1997),
and for dwarf irregular galaxies ({\it stars}; Garnett 1989). The curve
represents the sulfur ionization correction formula, equation [4],
assuming an intrinsic abundance ratio log\,(S/O) = $-1.6$.  }
\end{figure}

\clearpage 

\begin{figure}
\epsscale{0.9}
\plotone{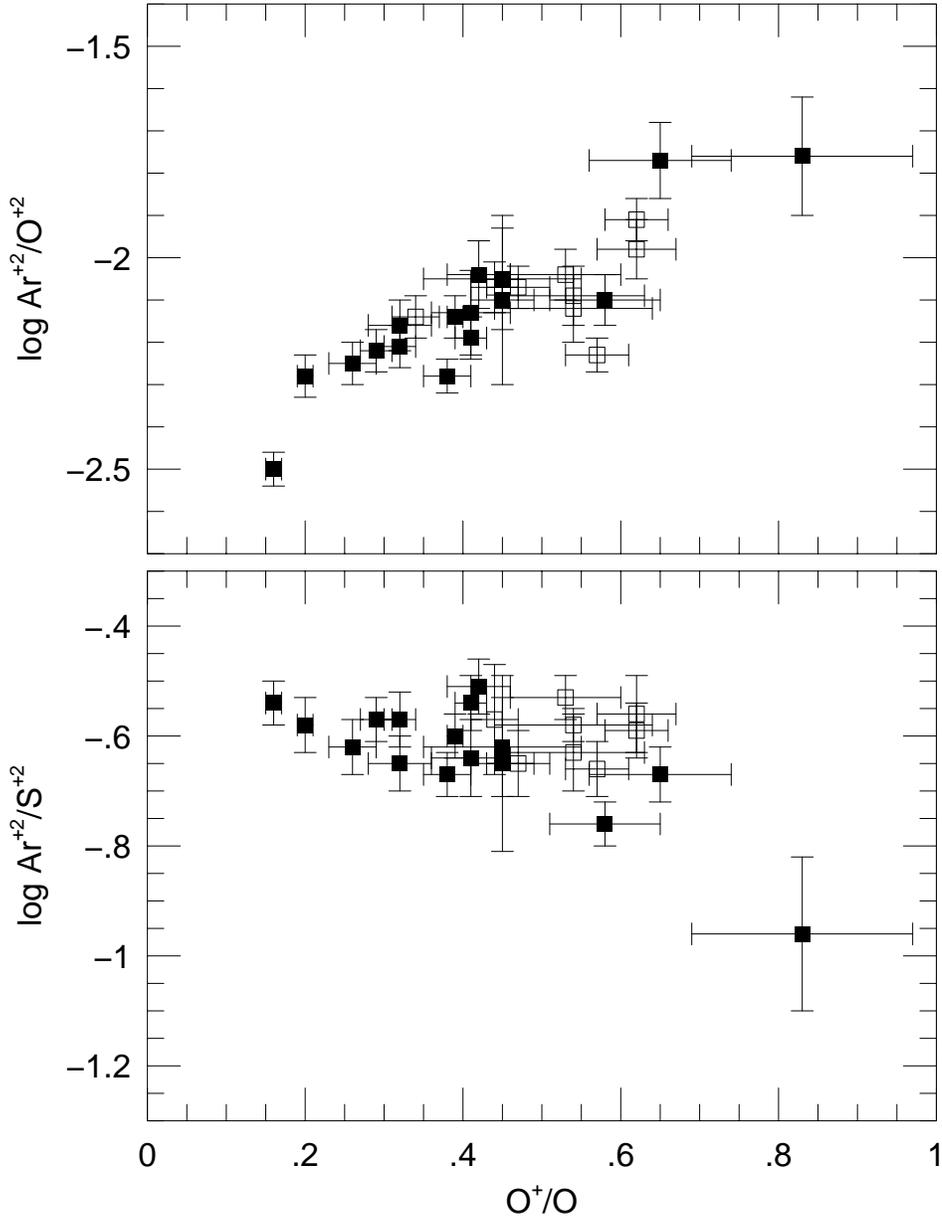}
\caption{Two comparisons showing the relative ionization behaviors of 
Ar, O, and S in extragalactic \hii\ regions.  
{\it Top}:  Ar$^{+2}$/O$^{+2}$ vs. O$^+$ ionic fraction.
{\it Bottom}:  Ar$^{+2}$/S$^{+2}$ vs. O$^+$ ionic fraction. Symbols are the
same as in Figures 2 and 5.  Except for H336 with O$^+$/O = 0.85, 
Ar$^{+2}$/S$^{+2}$
shows little correlation with ionization, in contrast to 
Ar$^{+2}$/O$^{+2}$. 
}
\end{figure}

\clearpage

\begin{figure}
\epsscale{1.0}
\plotone{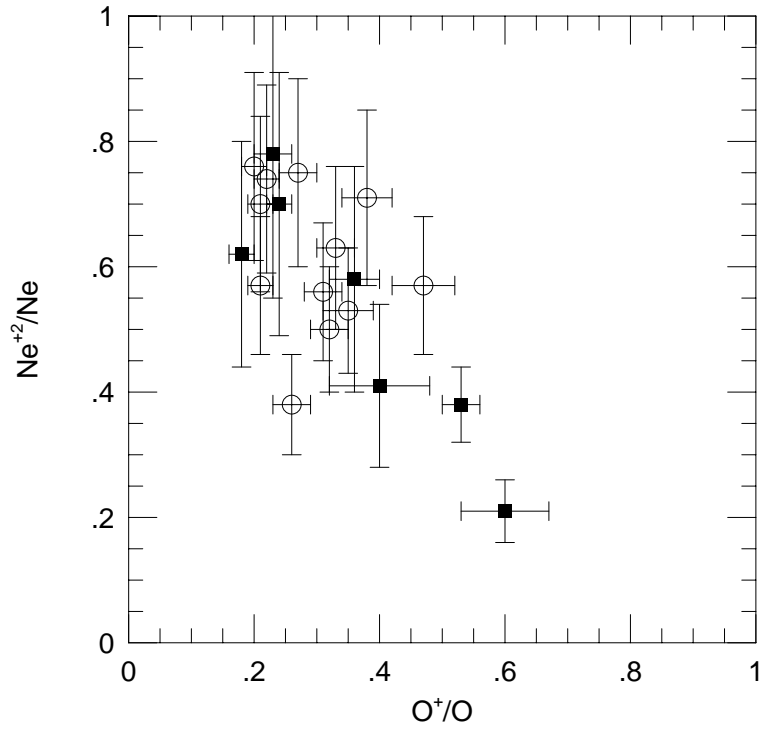}
\caption{The Ne$^{+2}$ ionic fraction as a function of O$^+$ ion
fraction, derived from optical and infrared spectroscopy of \hii\
regions. Open circles are LMC and SMC
regions from Vermeij \& van der Hulst (2002). Filled squares are M33
\hii\ regions from Willner \& Nelson-Patel (2002).  }
\end{figure}

\clearpage 

\begin{figure}
\plotone{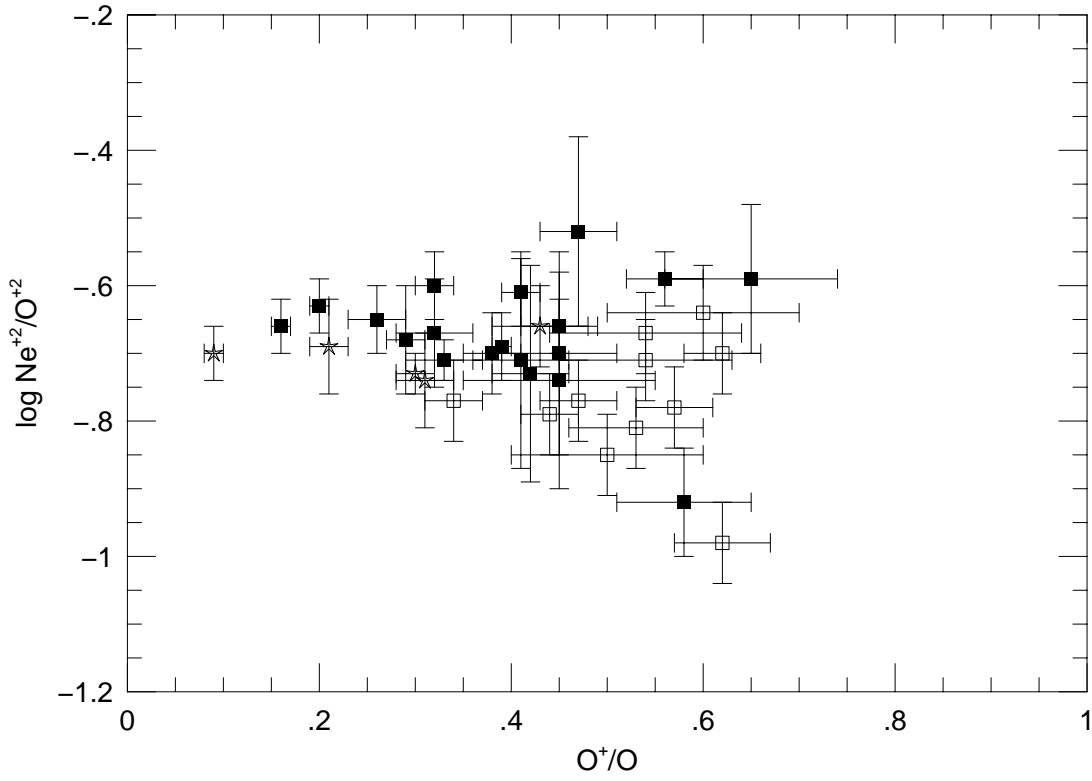}
\figcaption{ The Ne$^{+2}$/O$^{+2}$ ion abundance ratio for \hii\ regions in
M101 {(\it filled squares)} plotted as a function of O$^+$ fraction. Symbols
are the same as in Figure~2.
}
\end{figure}


\clearpage 

\begin{figure}
\plotone{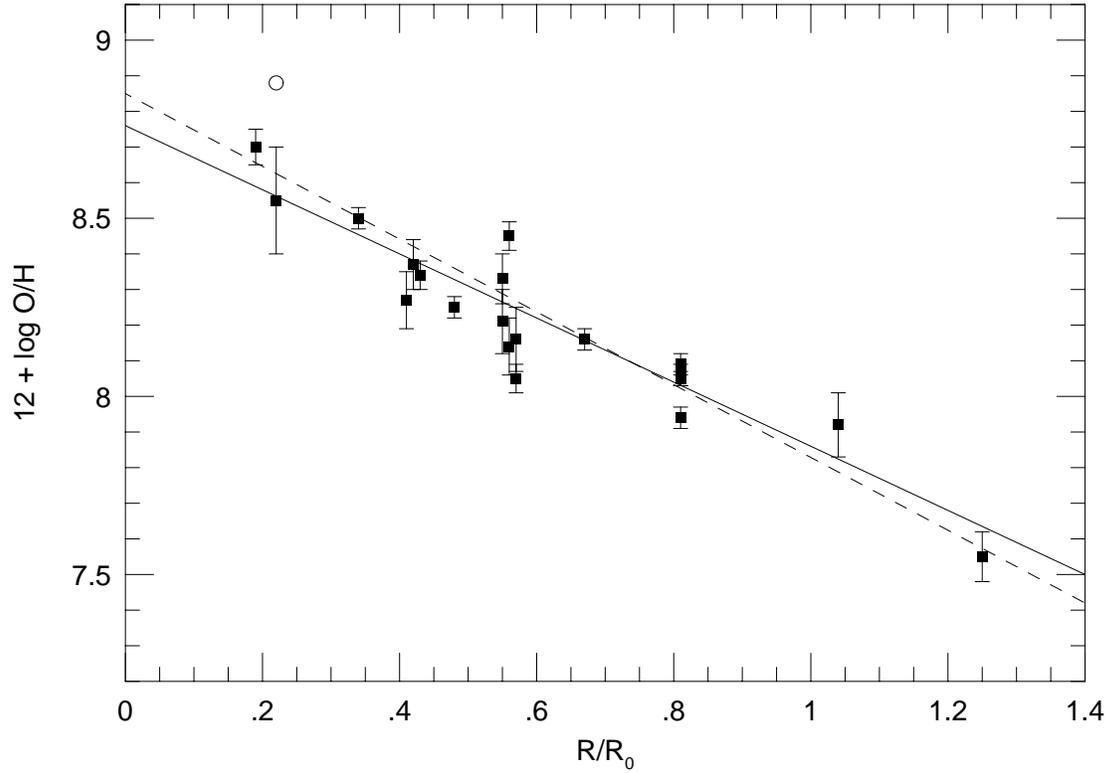}
\caption{ The oxygen abundance gradient in M101 from our 20 \hii\ regions 
with electron
temperature measurements. The linear fit to the data given by equation [5]
is shown by the solid line. The open circle is the abundance for H336 
obtained by Kinkel \& Rosa (1994). The dashed line shows the resulting 
fit if we use this point instead of our measurement, as given by equation [6].
}
\end{figure}

\clearpage

\begin{figure}
\plotone{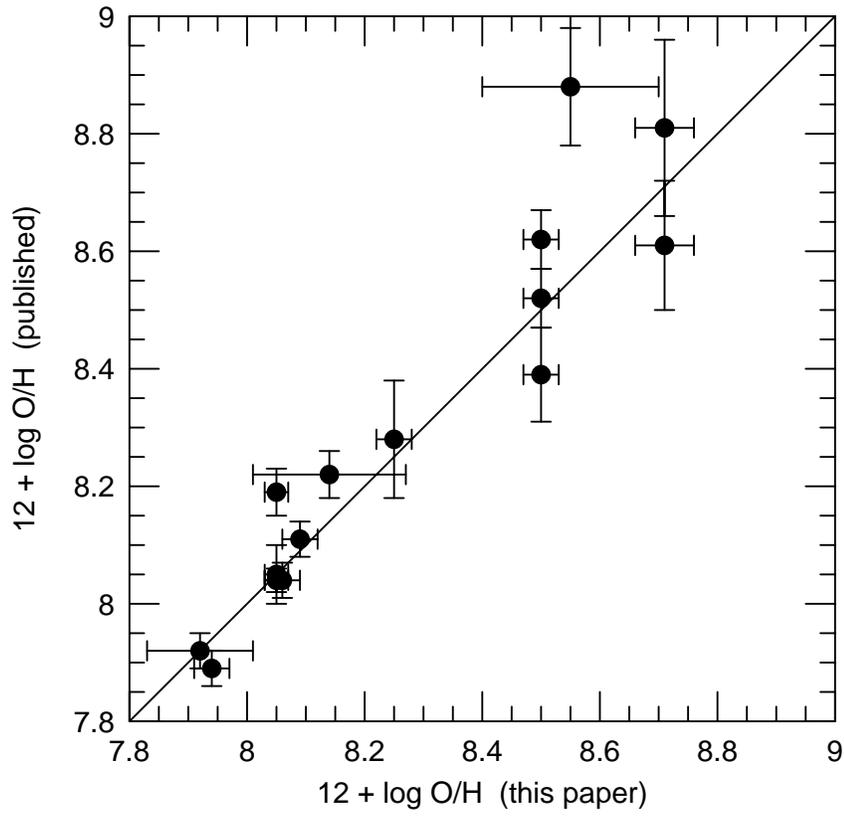}
\caption{ Comparison of our O/H abundance measurements for M101 \hii\ 
regions with those from the literature. Data are given in Table 6.
}
\end{figure}

\clearpage

\begin{figure}
\epsscale{0.9}
\plotone{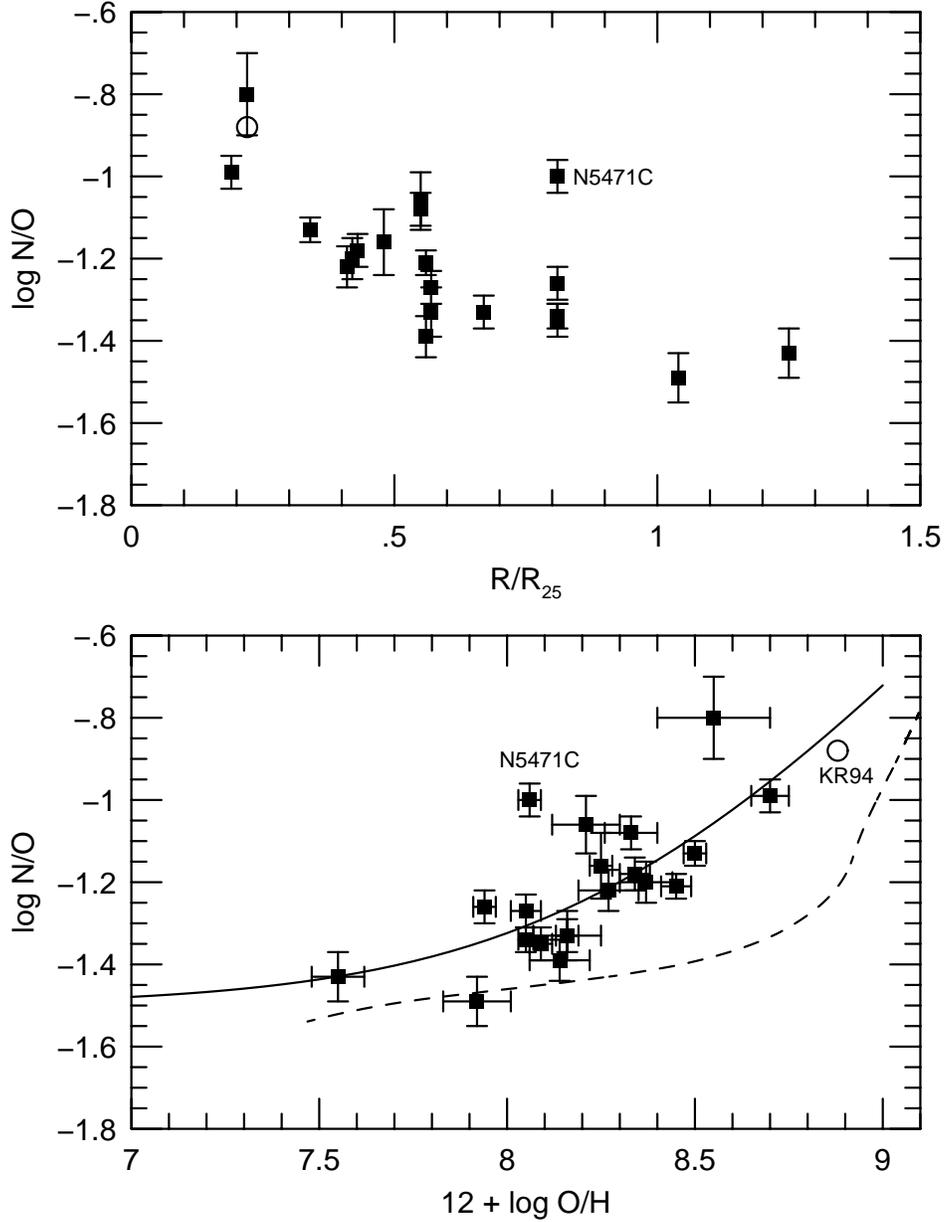}
\caption{{\it Top:} Radial variation of N/O for our M101 \hii\ region sample. 
{\it Bottom:} N/O vs. oxygen abundance.  The
open circle shows the abundances derived by Kinkel \& Rosa (1994) for H336.
The solid curve is a simple model for N/O which has a constant primary component
with log\,(N/O) = $-1.5$ and a secondary component in which N/O increases 
proportionally with
O/H. The dashed curve shows model B from Henry et al.~(2000), as discussed
in the text. } 
\end{figure}

\clearpage

\begin{figure}
\plotone{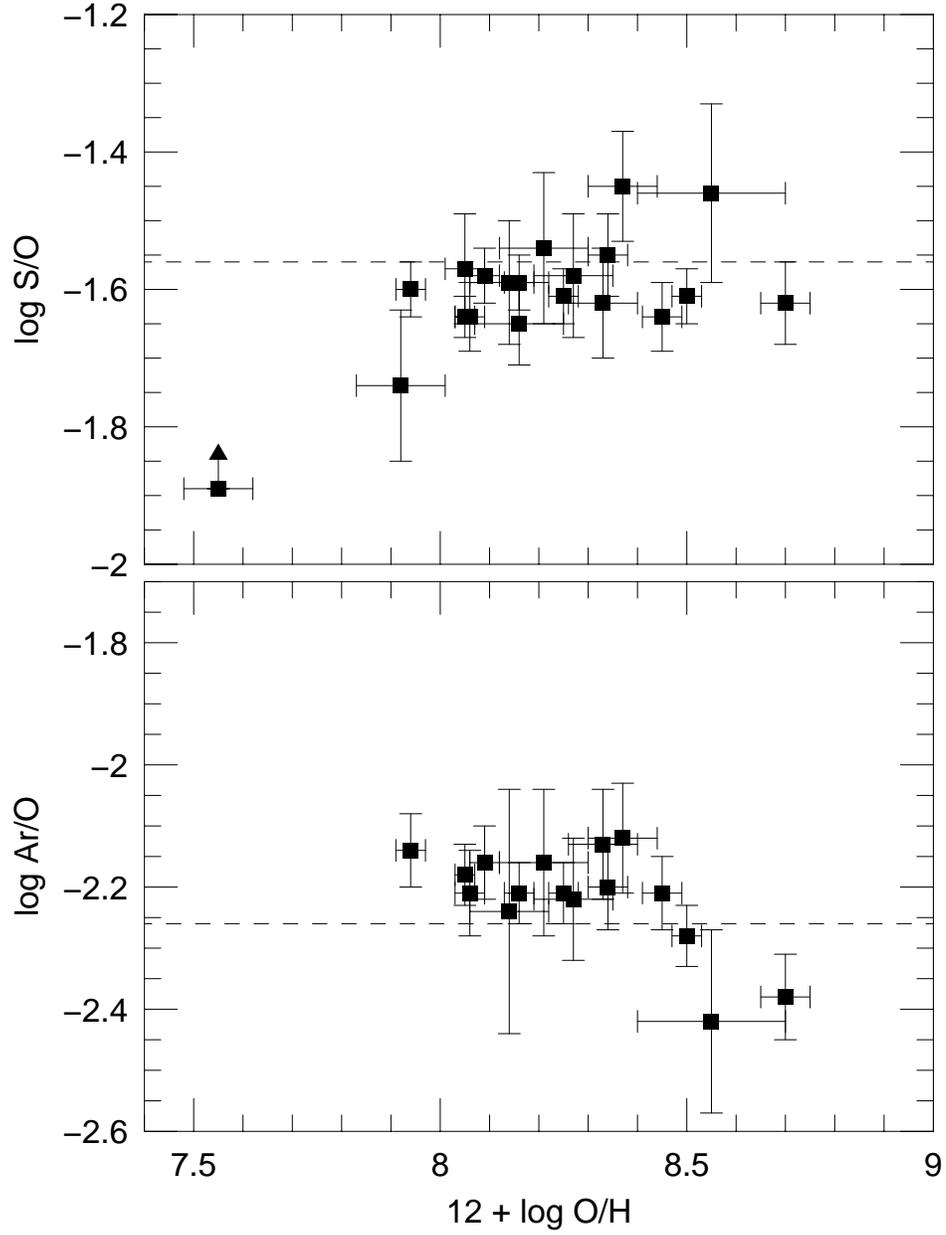}
\caption{Dependences of S/O ({\it Top}) and Ar/O ({\it Bottom}) on
oxygen abudance in M101.
The dashed lines show the average values for dwarf irregular
galaxies as derived by Izotov \& Thuan (1999).
}
\end{figure}

\clearpage

\begin{figure}
\epsscale{1.0}
\plotone{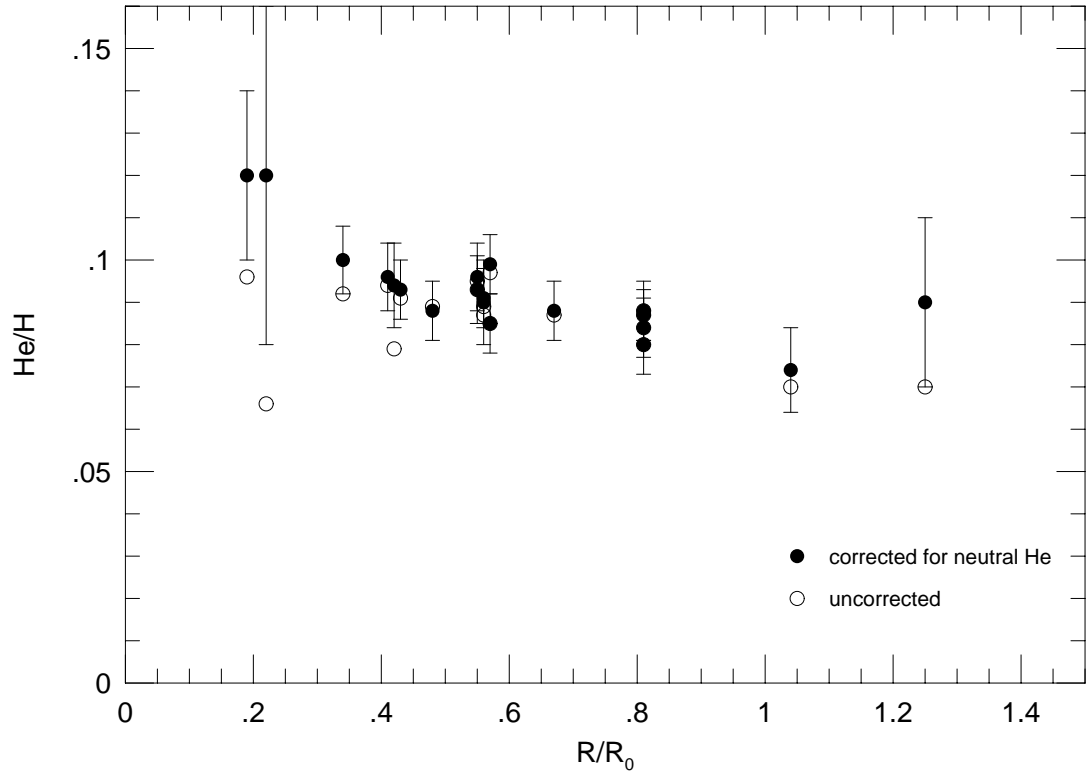}
\caption{Helium abundances in M101 plotted as a function of galactocentric
radius. The open circles show the uncorrected ionized He$^+$/H$^+$ ratios,
while the filled circles show the total He abundances, corrected for 
neutral helium as described in the text.
}
\end{figure}

\clearpage

\begin{figure}
\plotone{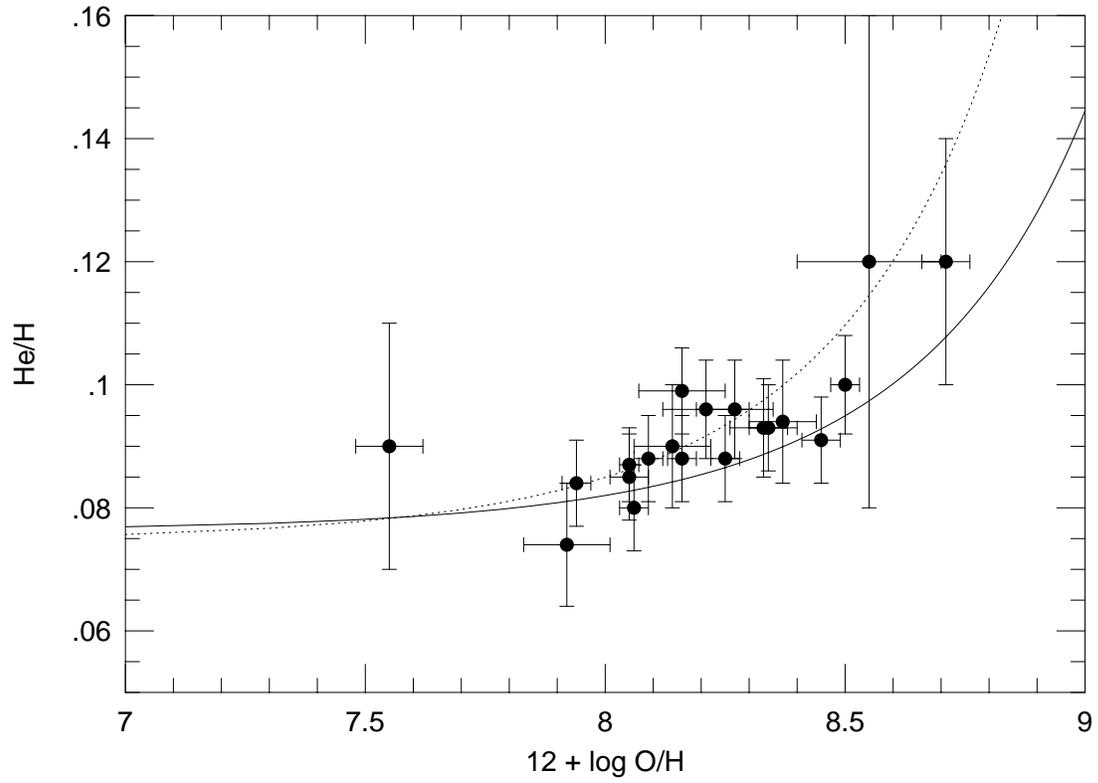}
\caption{ Helium abundances in M101 as a function of oxygen abundance.
The solid and dashed lines represent two relations derived from
measurements of metal-poor dwarf galaxies by Olive,
Skillman, \& Steigman (1998).
}
\end{figure}

\clearpage

\begin{figure}
\epsscale{0.8}
\plotone{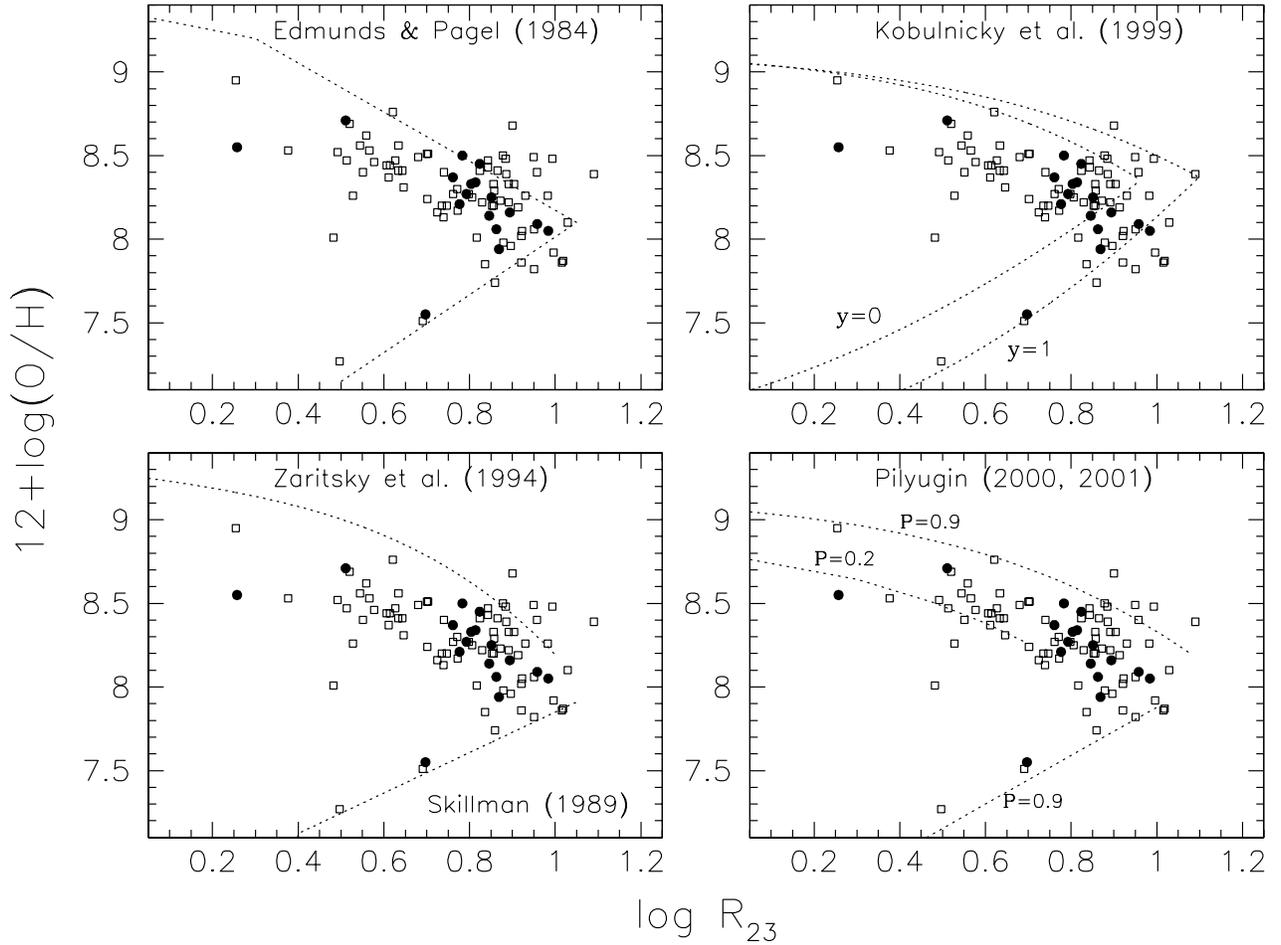}
\caption{Comparisons of different 'strong-line' empirical abundance
calibrations from the literature (dotted lines) with 'direct'
abundance determinations in the $\log R_{23}$ vs logarithmic oxygen
abundance plane.  Each panel contains the same data points; solid
circles denote M101 \hii\ regions from this paper, and open squares
show published measurements of other, mostly metal-rich regions (see
text for references).  The dashed lines in each panel show a different
$R_{23}$ calibration from the reference indicated.  The Kobulnicky et
al.\ (1999) and Pilyugin (2000, 2001) calibrations are parameterized
in terms of $R_{23}$ and a secondary measure of the hardness of the
radiation field ($y$ and $P$, respectively), and we have drawn curves
which bracket the typical values encountered in the \hii\ region
samples.}
\end{figure}

\clearpage

\begin{figure}
\epsscale{1.0}
\plotone{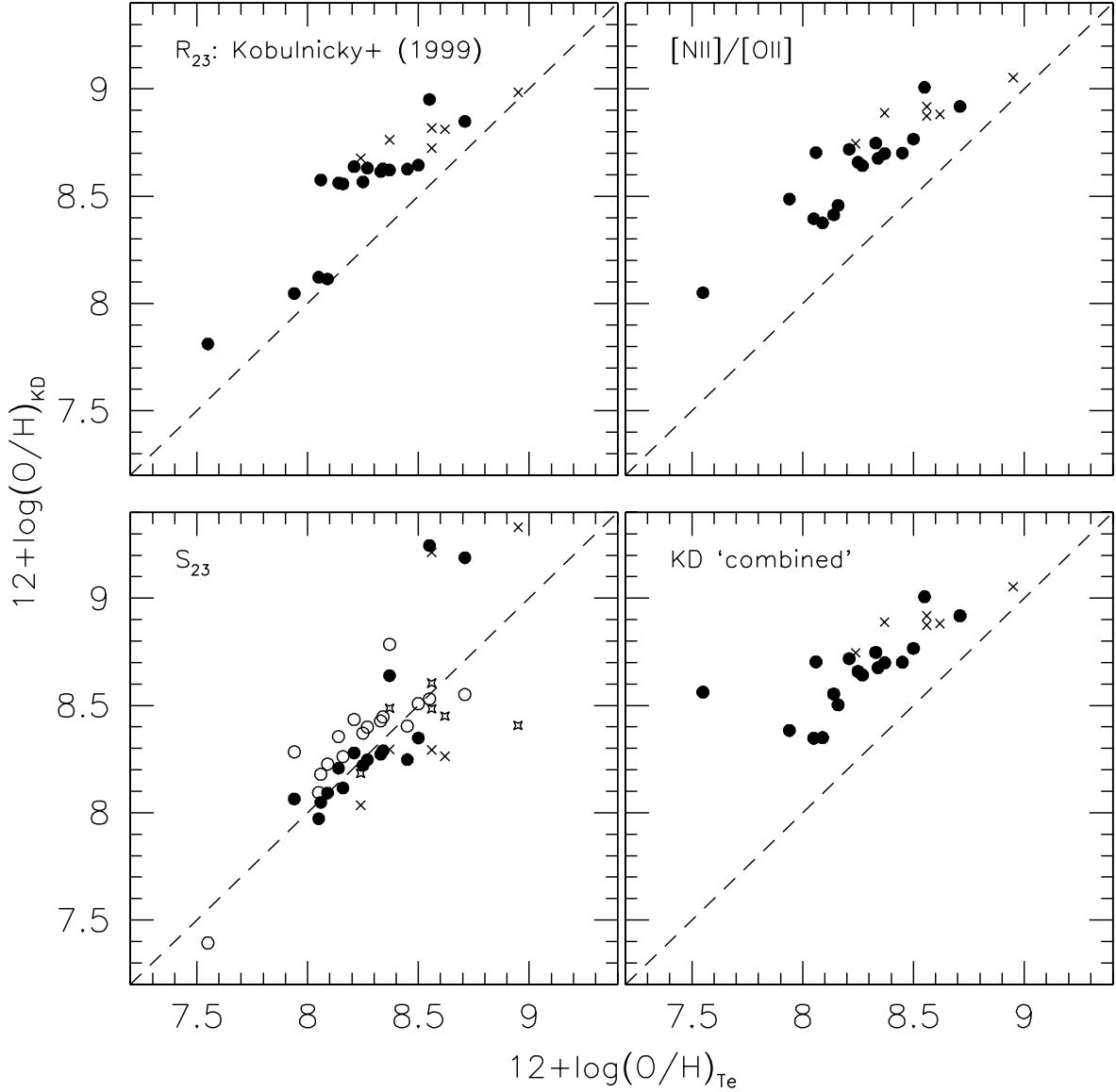}
\caption{Comparison of electron temperature based abundances
(O/H)$_{Te}$ with those derived from different strong-line
calibrations.  Full circles represent our M101 data, while crosses are
used for metal-rich \hii\ regions from Castellanos et al.~(2002) and
D\'\i az et al.~(2000).  The four panels show empirical abundances using
the calibrations of Kewley \& Dopita (2002) for $R_{23}$ (eq.~[7]),
[\nii]$\lambda$6584/[\oii]$\lambda\lambda$3726,3729, $S_{23} \equiv
{([\rm
\sii]\lambda\lambda6717,6731+[\siii]\lambda\lambda9069,9532)}$/$H\beta$,
and a combined index, as described in their paper.  For the
$S_{23}$ comparison we have also compared to the original D\'\i az \&
P\'erez-Montero (2000) calibration (open symbols).}
\end{figure}

\clearpage

\begin{figure}
\plotone{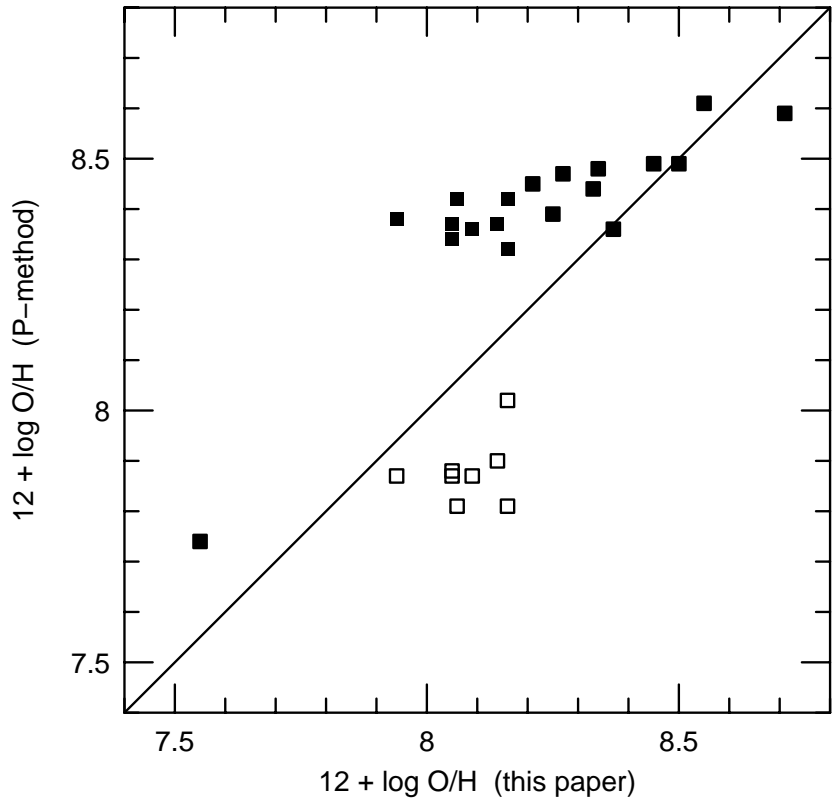}
\caption{Comparison of $T_e$-based abundances from this paper 
with those derived using
the Pilyugin (2001b) $P$-method, as described in the text.}
\end{figure}

\begin{deluxetable}{lcccccccccccccccccccccccc}
\setlength\tabcolsep{5pt}

\tabletypesize{\scriptsize}
\rotate
\tablewidth{0pt}
\tablenum{1}
\tablecaption{Dereddened Line Fluxes and Errors\tablenotemark{a}}

\tablehead{
\colhead{ID}            &
\multicolumn{2}{c}{[O\,II]}    & 
\multicolumn{2}{c}{[Ne\,III]}    & 
\multicolumn{2}{c}{[O\,III]}    & 
\multicolumn{2}{c}{[O\,III]}    & 
\multicolumn{2}{c}{[N\,II]}    & 
\multicolumn{2}{c}{[S\,III]}    & 
\multicolumn{2}{c}{[N\,II]}    & 
\multicolumn{2}{c}{[S\,II]}    & 
\multicolumn{2}{c}{[S\,II]}    & 
\multicolumn{2}{c}{[Ar\,III]}    & 
\multicolumn{2}{c}{[O\,II]}    & 
\multicolumn{2}{c}{[S\,III]}    \\

\colhead{}                      &
\multicolumn{2}{c}{3727}    & 
\multicolumn{2}{c}{3869}    & 
\multicolumn{2}{c}{4363}    & 
\multicolumn{2}{c}{5007}    & 
\multicolumn{2}{c}{5755}    & 
\multicolumn{2}{c}{6312}    & 
\multicolumn{2}{c}{6584}    & 
\multicolumn{2}{c}{6717}    & 
\multicolumn{2}{c}{6731}    & 
\multicolumn{2}{c}{7135}    & 
\multicolumn{2}{c}{7325}    & 
\multicolumn{2}{c}{9069+9532}    }
\startdata
\vspace{-3mm}
\\
H1013     & 188 &   10 &    3.0 &   0.2 &  \nodata & \nodata &  103 & 5 &    0.5 &   0.1 &    0.8 &   0.1 &   64.6 &   3.4 &   16.7 &   0.9 &   12.1 & 0.6 &  7.8 &   0.5 &    2.6 &   0.3 &  131.6 &   7.0 \\
H1105     & 185 &   10 &   18.3 &   1.0 &    1.4 &   0.1 &  316 &  17 &    0.4 & \nodata &    1.3 &   0.1 &   33.4 &   1.8 &   13.1 &   0.7 & 11.0 & 0.6 &   10.2 &   0.6 &    3.6 &   0.3 &  126.1 &   6.1 \\
H1159     & 198 &  10 &   20.1 &   1.2 &    1.9 &   0.4 &  317 &  17 &  \nodata & \nodata &    1.6 &   0.2 &   23.6 &   1.3 &   17.6 &  0.9 & 12.2 & 0.7 &   9.1 &   0.7 &    4.8 &   0.6 &   97.0 &   5.4 \\
H1170     & 308 &  16 &   15.8 &   0.9 &    1.6 &   0.2 &  201 &  11 &  \nodata & \nodata &    1.7 &   0.2 &   44.0 &   2.3 &   33.4 &  1.8 & 23.3 & 1.2 &   14.1 &   0.9 &    8.5 &   0.7 &  170.0 &   9.3 \\
H1176     & 160 &   8 &   25.0 &   1.3 &    2.4 &   0.3 &  369 &  20 &  \nodata & \nodata &    1.5 &   0.1 &   21.2 &   1.1 &   13.4 &  0.7 & 9.6 & 0.5  &   10.0 &   0.7 &    3.0 &   0.3 &  113.5 &   6.1 \\
H1216     & 151 &   8 &   36.1 &   1.9 &    4.7 &   0.3 &  473 &  25 &  \nodata & \nodata &    1.6 &   0.1 &   11.1 &   0.6 &   11.0 &  0.6 & 7.9 & 0.4  & 8.2 &   0.6 &    2.7 &   0.3 &   83.0 &   4.6 \\
H128      & 145 &   8 &   24.5 &   1.3 &    1.7 &   0.2 &  391 &  28 &    0.3 & \nodata &    1.3 &   0.1 &   20.8 &   1.1 &   13.3 &   0.7 &  10.0 & 0.5 & 10.6 &   0.7 &    4.1 &   0.3 &  104.4 &   5.7 \\
H140      & 317 &  17 &   10.0 &   1.2 &  \nodata & \nodata &  114 &   6 &  \nodata & \nodata &    1.9 &   1.5 &   60.2 &   3.4 &   48.3 &   3.1 & 32.6 & 2.3 &   5.2 &   1.0 &    9.1 &   1.2 &   75.3 &   6.0 \\
H143      & 219 &  12 &   17.1 &   1.0 &    2.3 &   0.4 &  284 &  15 &    0.9 &   0.2 &    1.4 &   0.2 &   33.3 &   1.8 &   23.2 &   1.2 &  17.0 & 0.9 &  9.2 &   0.7 &    2.4 &   0.4 &   93.8 &   5.2 \\
H149      & 212 &  11 &   18.4 &   1.0 &    1.8 &   0.4 &  318 &  17 &    0.9 &   0.2 &    1.4 &   0.1 &   35.9 &   1.9 &   21.1 &   1.1 &  16.1 & 0.8 & 11.6 &   0.8 &    5.6 &   0.4 &   95.4 &   5.0 \\
H203      & 223 &  12 &  \nodata & \nodata &  \nodata & \nodata &   44 &   2 &  \nodata & \nodata &  \nodata & \nodata &   91.1 &   4.9 &   38.7 &   2.2 & 27.9 & 1.8 &   5.3 &   0.5 &  \nodata & \nodata &   93.4 &   5.5 \\
H237      & 136 &   8 &    5.0 &   0.9 &  \nodata & \nodata &  114 &   6 &    3.2 &   1.3 &  \nodata & \nodata &   48.3 &   2.7 &   21.1 &  1.6 & 14.2 & 1.3 &   8.1 &   0.7 &    5.1 &   0.8 &   69.4 &   4.3 \\
H336      & 178 &   9 &  \nodata & \nodata &  \nodata & \nodata &   23 &   1 &    0.5 &   0.1 &    0.6 &   0.1 &   95.9 &   5.1 &   33.5 &  1.8 & 23.3 & 1.2 &   3.9 &   0.4 &    1.4 &   0.2 &  107.0 &   5.7 \\
H409      & 218 &  12 &   24.6 &   1.3 &    2.3 &   0.2 &  370 &  20 &    0.4 &   0.1 &    1.7 &   0.1 &   27.3 &   1.4 &   17.2 &   0.9 & 14.0 & 0.7 &   9.2 &   0.7 &    4.7 &   0.4 &   90.1 &   4.9 \\
H67       & 244 &  13 &   25.6 &   1.4 &    3.5 &   0.5 &  342 &  18 &  \nodata & \nodata &    1.8 &   0.4 &   16.3 &   0.9 &  15.6 &  1.1 & 10.7 & 0.8 &   8.7 &   0.8 &    5.4 &   0.8 &   92.1 &   5.7 \\
H70       & 311 &  16 &   23.6 &   1.4 &    2.5 &   0.5 &  267 &  14 &  \nodata & \nodata &    1.4 &   0.2 &   23.6 &   1.3 &   26.5 &   1.4 & 20.7 & 1.1 & \nodata & \nodata &  \nodata & \nodata &  \nodata & \nodata \\
H71       & 200 &  11 &   32.4 &   1.7 &    5.9 &   0.4 &  454 &  24 &  \nodata & \nodata &    1.9 &   0.3 &   15.9 &   0.9 &  17.2  &   1.0 & 11.9 & 0.7 & \nodata & \nodata &  \nodata & \nodata &  \nodata & \nodata \\
H875      & 283 &  15 &  \nodata & \nodata &  \nodata & \nodata &  102 &   5 &  \nodata & \nodata &  \nodata & \nodata &   64.1 &   3.4 &   36.4 &   1.9 & 26.3 & 1.4 &  6.5 &   0.6 &    3.6 &   0.4 &   75.3 &   4.8 \\
H972      & 148 &   8 &  \nodata & \nodata &  \nodata & \nodata &   30 &   2 &  \nodata & \nodata &  \nodata & \nodata &   86.9 &   4.7 &   22.3 &   1.4 & 16.1 & 1.1 &   3.8 &   0.5 &  \nodata & \nodata &   98.3 &   5.3 \\
H974      & 166 &  10 &  \nodata & \nodata &  \nodata & \nodata &   28 &   3 &  \nodata & \nodata &  \nodata & \nodata &   73.6 &   4.9 &   33.1 &   3.9 & 23.5 & 3.6 & \nodata & \nodata &  \nodata & \nodata &   37.6 &  11.8 \\
N5471-A   & 106 &   6 &   52.2 &   2.8 &    9.5 &   0.5 &  644 &  34 &  \nodata & \nodata &    1.6 &   0.1 &    6.8 &   0.4 &   8.7 &   0.5 & 7.1 & 0.4 &   7.7 &   0.6 &    3.8 &   0.3 &   61.4 &   3.4 \\
N5471-B   & 213 &  11 &   35.4 &   1.9 &    6.6 &   0.4 &  395 &  21 &    0.4 &   0.1 &    1.3 &   0.1 &   16.7 &   0.9 &   29.1 &   1.5 & 25.6 & 1.3 &   6.4 &   0.4 &    8.6 &   0.6 &   50.9 &   3.1 \\
N5471-C   & 174 &   9 &   34.4 &   1.8 &    5.4 &   0.3 &  416 &  22 &    1.2 &   0.1 &    1.4 &   0.1 &   25.1 &   1.3 &   13.1 &   0.7 & 10.1 & 0.5 &  6.8 &   0.5 &    5.2 &   0.4 &   65.9 &   3.7 \\
N5471-D   & 137 &   7 &   48.4 &   2.6 &    8.0 &   0.4 &  578 &  31 &  \nodata & \nodata &    1.7 &   0.1 &    8.5 &   0.5 &   11.7 &   0.6 & 8.9 & 0.5 &  8.0 &   0.6 &    1.8 &   0.3 &   75.7 &   4.3 \\
SDH 323    & 194 &  10 &   28.0 &   1.7 &    5.5 &   0.9 &  227 &  12 &  \nodata & \nodata &  \nodata & \nodata &    7.9 &   0.7 &  12.0 &  1.0 & 8.8 & 1.0 & \nodata & \nodata &  \nodata & \nodata &  \nodata & \nodata \\
\enddata
\tablenotetext{a}{The first column in each pair of values gives 
line fluxes normalized to $H\beta = 100$, after 
corrections for reddening and stellar absorption have been applied.
The second column in each pair gives the respective uncertainty, as
described in the text.}
\end{deluxetable}

\clearpage

\begin{deluxetable}{lcccccccc}
\tabletypesize{\scriptsize}
\tablecolumns{9}
\tablewidth{0pt}
\tablenum{2}
\tablecaption{Measured Electron Temperatures (K)}

\tablehead{
\colhead{ID}            &
\multicolumn{2}{c}{T[O\,II]\tablenotemark{a}}	&
\multicolumn{2}{c}{T[N\,II]\tablenotemark{a}}	&
\multicolumn{2}{c}{T[S\,III]\tablenotemark{a}}	&
\multicolumn{2}{c}{T[O\,III]\tablenotemark{a}}	}
\startdata
\vspace{-3mm}
\\
H1013      &    9000 &     400 &    8100 &     300 &    7200 &     200 & \nodata & \nodata \\
H1105      &    9300 &     400 &    9400 &     400 &    8500 &     200 &    8900 &     200 \\
H1159      &    8100 &     400 & \nodata & \nodata &   10200 &     500 &    9800 &     600 \\
H1170      &    8600 &     300 & \nodata & \nodata &    8500 &     300 &   10800 &     400 \\
H1176      &   10500 &     600 & \nodata & \nodata &    9300 &     300 &   10000 &     300 \\
H1216      &   10300 &     600 & \nodata & \nodata &   11000 &     400 &   11500 &     200 \\
H128       &   13000 &     800 &   10700 &     600 &    9200 &     200 &    8900 &     300 \\
H143       &    8100 &     600 &   14100 &    2200 &    9800 &     600 &   10800 &     700 \\
H149       &   12100 &     600 &   13600 &    1300 &    9800 &     300 &    9600 &     600 \\
H336       &    7200 &     300 &    7300 &     400 &    7000 &     500 & \nodata & \nodata \\
H409       &   10100 &     500 &    9900 &    1000 &   10900 &     300 &   10000 &     200 \\
H67        &    7800 &     500 & \nodata & \nodata &   11100 &    1300 &   11600 &     600 \\
H70        & \nodata & \nodata & \nodata & \nodata & \nodata & \nodata &   11300 &     800 \\
H71        & \nodata & \nodata & \nodata & \nodata & \nodata & \nodata &   12700 &     400 \\
N5471-A    &   13800 &     800 & \nodata & \nodata &   12800 &     400 &   13400 &     300 \\
N5471-B    &   13900 &     700 &   13500 &    1200 &   12700 &     500 &   14000 &     300 \\
N5471-C    &   13200 &     700 &   20000 &    1800 &   11500 &     500 &   12700 &     300 \\
N5471-D    &    8400 &     600 & \nodata & \nodata &   11800 &     500 &   13000 &     300 \\
SDH323    & \nodata & \nodata & \nodata & \nodata & \nodata & \nodata &   16700 &    1400 \\
\enddata
\tablenotetext{a}{The left column gives the observed electron temperature,
and the right column gives the corresponding uncertainty.}
\end{deluxetable}

\begin{deluxetable}{lcccccc}
\tabletypesize{\scriptsize}
\tablecolumns{7}
\tablewidth{0pt}
\tablenum{3}
\tablecaption{Adopted Electron Temperatures (K)}

\tablehead{
\colhead{ID}            &
\multicolumn{2}{c}{T(O$^+$,N$^+$,S$^+$)\tablenotemark{a}}	&
\multicolumn{2}{c}{T(S$^{+2}$,Ar$^{+2}$)\tablenotemark{a}}      &
\multicolumn{2}{c}{T(O$^{+2}$, Ne$^{+2}$)\tablenotemark{a}}	}
\startdata
\vspace{-3mm}
\\
H1013      &    7600 &     500 &    7200 &     200 &    6600 &     500 \\
H1105      &    9000 &     500 &    8800 &     200 &    8600 &     200 \\
H1159      &   10000 &     800 &   10000 &     500 &   10000 &     600 \\
H1170      &    9600 &     600 &    9600 &     300 &    9500 &     400 \\
H1176      &    9700 &     600 &    9600 &     300 &    9600 &     300 \\
H1216      &   11000 &     600 &   11100 &     400 &   11400 &     200 \\
H128       &    9300 &     600 &    9200 &     200 &    9000 &     300 \\
H143       &   10200 &     800 &   10200 &     600 &   10300 &     700 \\
H149       &    9800 &     700 &    9800 &     300 &    9700 &     600 \\
H336       &    7400 &     700 &    6900 &     500 &    6300 &     500 \\
H409       &   10400 &     600 &   10400 &     300 &   10600 &     200 \\
H67        &   11000 &     800 &   11200 &    1300 &   11400 &     600 \\
H70        &   10900 &     900 &   11100 &     900 &   11300 &     800 \\
H71        &   11900 &     600 &   12200 &     600 &   12700 &     400 \\
N5471-A    &   12400 &     600 &   12800 &     400 &   13400 &     300 \\
N5471-B    &   12500 &     600 &   13000 &     500 &   13300 &     300 \\
N5471-C    &   11500 &     700 &   11800 &     500 &   12200 &     300 \\
N5471-D    &   11800 &     600 &   12200 &     500 &   12600 &     300 \\
SDH 323    &   14700 &    1500 &   15600 &    1500 &   16700 &    1400 \\
\enddata
\tablenotetext{a}{Adopted electron temperatures for the low, moderate, and 
high-ionization zones, as described in \S~3.2.  The second column in
each case lists the corresponding uncertainty in $T_e$}
\end{deluxetable}

\begin{deluxetable}{lcccccccccccccc}
\tabletypesize{\scriptsize}
\tablecolumns{15}
\tablewidth{0pt}
\tablenum{4}
\tablecaption{Ionic Abundances: 12 + log\,(X/H)\tablenotemark{a}}

\tablehead{
\colhead{ID}            &
\multicolumn{2}{c}{O$^+$}	&
\multicolumn{2}{c}{O$^{++}$}	&
\multicolumn{2}{c}{S$^+$}	&
\multicolumn{2}{c}{S$^{++}$}	&
\multicolumn{2}{c}{N$^+$}	&
\multicolumn{2}{c}{Ne$^{++}$}	&
\multicolumn{2}{c}{Ar$^{++}$}	}
\startdata
\vspace{-3mm}
\\
 H1013    &  8.46 &  0.07 &  8.34 &  0.05 &  6.14 &  0.03 &  7.00 &  0.03 &  7.47 &  0.04 &  7.42 &  0.06 &  6.29 &  0.03 \\
 H1105    &  8.08 &  0.05 &  8.29 &  0.04 &  5.88 &  0.03 &  6.78 &  0.02 &  6.95 &  0.03 &  7.59 &  0.05 &  6.18 &  0.02 \\
 H1159    &  7.88 &  0.10 &  8.05 &  0.10 &  5.84 &  0.06 &  6.56 &  0.05 &  6.66 &  0.06 &  7.34 &  0.12 &  5.91 &  0.05 \\
 H1170    &  8.18 &  0.08 &  7.93 &  0.06 &  6.15 &  0.04 &  6.83 &  0.03 &  6.98 &  0.04 &  7.34 &  0.07 &  6.32 &  0.04 \\
 H1176    &  7.84 &  0.07 &  8.17 &  0.05 &  5.75 &  0.04 &  6.66 &  0.03 &  6.66 &  0.04 &  7.50 &  0.06 &  6.05 &  0.04 \\
 H1216    &  7.57 &  0.07 &  8.03 &  0.03 &  5.54 &  0.04 &  6.40 &  0.03 &  6.24 &  0.04 &  7.38 &  0.03 &  5.81 &  0.04 \\
 H128     &  7.91 &  0.05 &  8.30 &  0.05 &  5.81 &  0.03 &  6.65 &  0.02 &  6.70 &  0.03 &  7.62 &  0.06 &  6.09 &  0.03 \\
 H143     &  7.86 &  0.13 &  7.96 &  0.10 &  5.95 &  0.07 &  6.53 &  0.06 &  6.80 &  0.08 &  7.22 &  0.11 &  5.96 &  0.07 \\
 H149     &  7.95 &  0.07 &  8.09 &  0.10 &  5.95 &  0.04 &  6.56 &  0.03 &  6.87 &  0.04 &  7.36 &  0.12 &  6.08 &  0.04 \\
 H336     &  8.47 &  0.14 &  7.76 &  0.16 &  6.50 &  0.11 &  6.96 &  0.09 &  7.67 &  0.12 & \nodata & \nodata &  6.00 &  0.11 \\
 H409     &  7.84 &  0.06 &  8.03 &  0.03 &  5.83 &  0.03 &  6.49 &  0.03 &  6.68 &  0.04 &  7.34 &  0.04 &  5.87 &  0.03 \\
 H67      &  7.79 &  0.11 &  7.89 &  0.07 &  5.69 &  0.05 &  6.44 &  0.11 &  6.40 &  0.07 &  7.23 &  0.08 &  5.81 &  0.12 \\
 H70      &  7.91 &  0.13 &  7.81 &  0.09 &  5.96 &  0.08 &  6.34 &  0.14 &  6.58 &  0.09 &  7.23 &  0.10 & \nodata & \nodata \\
 H71      &  7.57 &  0.08 &  7.88 &  0.04 &  5.63 &  0.05 &  6.32 &  0.07 &  6.30 &  0.06 &  7.18 &  0.04 & \nodata & \nodata \\
 N5471-A  &  7.25 &  0.05 &  7.97 &  0.03 &  5.37 &  0.03 &  6.16 &  0.02 &  5.91 &  0.03 &  7.31 &  0.03 &  5.63 &  0.03 \\
 N5471-B  &  7.55 &  0.07 &  7.72 &  0.03 &  5.90 &  0.04 &  6.07 &  0.03 &  6.29 &  0.04 &  7.11 &  0.03 &  5.57 &  0.04 \\
 N5471-C  &  7.57 &  0.07 &  7.89 &  0.03 &  5.59 &  0.04 &  6.25 &  0.03 &  6.57 &  0.05 &  7.29 &  0.03 &  5.66 &  0.04 \\
 N5471-D  &  7.39 &  0.07 &  7.99 &  0.03 &  5.52 &  0.04 &  6.29 &  0.03 &  6.05 &  0.04 &  7.36 &  0.03 &  5.70 &  0.04 \\
 SDH323   &  7.22 &  0.11 &  7.27 &  0.09 &  5.33 &  0.06 & \nodata & \nodata &  5.79 &  0.07 &  6.75 &  0.10 & \nodata & \nodata \\
\enddata
\tablenotetext{a}{The first column in each pair lists the abundance, and 
the second column lists the corresponding (logarithmic) uncertainty.}
\end{deluxetable}

\begin{deluxetable}{lccccccccccc}
\tabletypesize{\scriptsize}
\tablecolumns{11}
\tablewidth{0pt}
\tablenum{5}
\tablecaption{Total abundances\tablenotemark{a}}

\tablehead{
\colhead{ID}            &
\colhead{R/R$_0$}       &
\multicolumn{2}{c}{12+log\,(O/H)}	&
\multicolumn{2}{c}{log\,(N/O)}	&
\multicolumn{2}{c}{log\,(S/O)}	&
\multicolumn{2}{c}{log\,(Ar/O)}	&
\multicolumn{2}{c}{log\,(Ne/O)}	}
\startdata
\vspace{-3mm}
\\
 H1013   & 0.19 & 8.71 & 0.05 & --0.99 & 0.03 & --1.62 & 0.06 & --2.38 & 0.07 & --0.92 & 0.08 \\
 H336    & 0.22 & 8.55 & 0.16 & --0.80 & 0.09 & --1.46 & 0.16 & --2.42 & 0.07 & \nodata & \nodata \\
 H1105   & 0.34 & 8.50 & 0.03 & --1.13 & 0.02 & --1.56 & 0.04 & --2.28 & 0.07 & --0.70 & 0.06 \\
 H1159   & 0.41 & 8.27 & 0.08 & --1.22 & 0.04 & --1.54 & 0.09 & --2.22 & 0.07 & --0.71 & 0.16 \\
 H1170   & 0.42 & 8.37 & 0.07 & --1.20 & 0.04 & --1.44 & 0.08 & --2.12 & 0.07 & --0.59 & 0.11 \\
 H1176   & 0.43 & 8.34 & 0.04 & --1.18 & 0.03 & --1.48 & 0.06 & --2.20 & 0.07 & --0.67 & 0.08 \\
 H409    & 0.48 & 8.25 & 0.03 & --1.16 & 0.07 & --1.61 & 0.04 & --2.21 & 0.07 & --0.69 & 0.05 \\
 H143    & 0.55 & 8.21 & 0.13 & --1.06 & 0.06 & --1.48 & 0.11 & --2.16 & 0.07 & --0.74 & 0.16 \\
 H149    & 0.55 & 8.33 & 0.07 & --1.08 & 0.03 & --1.58 & 0.08 & --2.13 & 0.07 & --0.73 & 0.16 \\
 H128    & 0.56 & 8.45 & 0.04 & --1.21 & 0.02 & --1.60 & 0.05 & --2.21 & 0.07 & --0.68 & 0.08 \\
 H67     & 0.56 & 8.14 & 0.08 & --1.39 & 0.05 & --1.52 & 0.05 & --2.24 & 0.07 & --0.66 & 0.04 \\
 H70     & 0.57 & 8.16 & 0.09 & --1.33 & 0.06 & --1.65 & 0.06 & \nodata & \nodata & --0.58 & 0.04 \\
 H71     & 0.57 & 8.05 & 0.04 & --1.27 & 0.04 & --1.57 & 0.08 & \nodata & \nodata & --0.71 & 0.03 \\
 H1216   & 0.67 & 8.16 & 0.03 & --1.33 & 0.03 & --1.54 & 0.04 & --2.21 & 0.07 & --0.65 & 0.05 \\
 N5471-A & 0.81 & 8.05 & 0.02 & --1.34 & 0.02 & --1.57 & 0.03 & --2.18 & 0.07 & --0.66 & 0.04 \\
 N5471-B & 0.81 & 7.94 & 0.03 & --1.26 & 0.03 & --1.53 & 0.04 & --2.14 & 0.07 & --0.61 & 0.05 \\
 N5471-C & 0.81 & 8.06 & 0.03 & --1.00 & 0.03 & --1.55 & 0.05 & --2.21 & 0.07 & --0.60 & 0.05 \\
 N5471-D & 0.81 & 8.09 & 0.03 & --1.35 & 0.03 & --1.50 & 0.04 & --2.16 & 0.07 & --0.63 & 0.04 \\
 H681\tablenotemark{b} & 1.04 & 7.92 & 0.09 & --1.49 & 0.14 & --1.74 & 0.12 & \nodata & \nodata & --0.70 & 0.15 \\
 SDH 323  & 1.25 & 7.55 & 0.07 & --1.43 & 0.05 & $>$ --1.89 & 0.11 & \nodata & \nodata & --0.52 & 0.14 \\
\enddata
\tablenotetext{a}{The first column in each pair lists the abundance, and 
the second column lists the corresponding (logarithmic) uncertainty.}
\tablenotetext{b}{Garnett \& Kennicutt (1994)}
\end{deluxetable}

\clearpage

\begin{deluxetable}{lccccccc}
\tabletypesize{\scriptsize}
\tablecolumns{8}
\tablewidth{0pt}
\tablenum{6}
\tablecaption{Comparison of 12 + log\,(O/H) with Published Results\tablenotemark{a}}

\tablehead{ \colhead{ID}            &
\colhead{this study }                      &
\colhead{RPT82}                      &
\colhead{TPF89}                      &
\colhead{Z98}                      &
\colhead{S85}                      &
\colhead{MRS85}                      &
\colhead{KR94}                      
}
\startdata
\vspace{-3mm}
\\
H1013     & 8.71$\pm$0.15 & 8.81$\pm$0.15 & 8.61$\pm$0.11 & \nodata       & \nodata       & \nodata       & \nodata \\
H1105     & 8.50$\pm$0.03 & 8.62$\pm$0.05 & 8.39$\pm$0.08 & \nodata       &  \nodata      & 8.52$\pm$0.05 & \nodata  \\
H336      & 8.55$\pm$0.15 &  \nodata      &  \nodata      & \nodata       &  \nodata      &  \nodata      & 8.88$\pm$00.10 \\
H409      & 8.25$\pm$0.03 & \nodata       & 8.28$\pm$0.10 & \nodata       &  \nodata      &  \nodata      &  \nodata  \\
H67       & 8.14$\pm$0.13 & \nodata       &  \nodata      & 8.22$\pm$0.04 &  \nodata      &  \nodata      &  \nodata  \\
NGC5471-A & 8.05$\pm$0.02 & 8.19$\pm$0.04 & 8.05$\pm$0.05 & \nodata       & 8.04$\pm$0.02 &  \nodata      &  \nodata  \\
NGC5471-B & 7.94$\pm$0.03 & \nodata       &  \nodata      & \nodata       & 7.89$\pm$0.03 &  \nodata      &  \nodata  \\
NGC5471-C & 8.06$\pm$0.03 & \nodata       &  \nodata      & \nodata       & 8.04$\pm$0.03 &  \nodata      &  \nodata  \\
NGC5471-D & 8.09$\pm$0.03 & \nodata       &  \nodata      & \nodata       & 8.11$\pm$0.03 &  \nodata      &  \nodata  \\
H681      & 7.92$\pm$0.09 & \nodata       &  \nodata      & 7.92$\pm$0.03 &  \nodata      &  \nodata      &  \nodata  \\
\enddata
\tablenotetext{a}{Sources for data in Table 8: RPT82: Rayo et al.~1982; 
TPF89: Torres-Peimbert et al.~1989; Z98: van Zee et al.~1998; S85: 
Skillman 1985; MRS85: McCall, Rybski, \& Shields 1985; KR94: Kinkel \& Rosa 1994.}
\end{deluxetable}

\clearpage

\begin{deluxetable}{lcccccccccccc}
\tabletypesize{\scriptsize}
\tablecolumns{13}
\tablewidth{0pt}
\tablenum{7}
\tablecaption{He I Line Fluxes and Equivalent Widths}

\tablehead{
\colhead{ID}      &
\colhead{I($\lambda$4026)\tablenotemark{a}}    &
\colhead{err.}     &
\colhead{EW (\AA)}      &
\colhead{I($\lambda$4471)\tablenotemark{a}}    &
\colhead{err.}     &
\colhead{EW (\AA)}      &
\colhead{I($\lambda$5876)\tablenotemark{a}}    &
\colhead{err.}     &
\colhead{EW (\AA)}      &
\colhead{I($\lambda$6678)\tablenotemark{a}}    &
\colhead{err.}     &
\colhead{EW}                      
}
\startdata
\vspace{-3mm}
\\
H1013   & 1.7 & 0.2 & 1 & 3.9 & 0.2 & 5 & 12.2 & 0.7 & 35 & 4.5 & 0.3 & 14 \\
H1105   & 1.9 & 0.1 & 2 & 4.2 & 0.2 & 7 & 12.6 & 0.7 & 46 & 3.6 & 0.2 & 17 \\
H1159   & 1.9 & 0.4 & 1 & 4.3 & 0.4 & 5 & 12.5 & 0.7 & 28 & 3.5 & 0.2 & 10 \\
H1170   & 1.3 & 0.2 & 1 & 3.4 & 0.3 & 3 & 10.3 & 0.6 & 22 & 2.7 & 0.2 & \ 8 \\
H1176   & 2.2 & 0.2 & 3 & 3.8 & 0.3 & 8 & 11.7 & 0.6 & 59 & 3.5 & 0.2 & 23 \\
H1216   & 1.4 & 0.2 & 1 & 3.8 & 0.2 & 5 & 10.7 & 0.6 & 27 & 3.4 & 0.2 & 12 \\
H128    & 1.7 & 0.2 & 2 & 3.9 & 0.2 & 5 & 12.7 & 0.7 & 38 & 3.5 & 0.2 & 14 \\
H140    & \nodata & \nodata & \nodata & \nodata & \nodata & \nodata & 11.4 & 1.4 & 13 & \nodata & \nodata & \nodata\\
H143    & \nodata & \nodata & \nodata & 3.5 & 0.5 & 2 & 12.6 & 0.7 & 22 & 3.6 & 0.3 & \ 9 \\
H149    & 2.0 & 0.3 & 4 & 4.1 & 0.3 & 9 & 12.7 & 0.7 & 63 & 3.6 & 0.2 & 24 \\
H203    & \nodata & \nodata & \nodata & \nodata & \nodata & \nodata & 10.4 & 0.8 & 24 & \nodata & \nodata & \nodata \\
H237    & \nodata & \nodata & \nodata & 4.5 & 0.8 & 2 & 12.1 & 1.0 & 11 & 2.9 & 0.7 & \ 4 \\
H336    & \nodata & \nodata & \nodata & 2.3 & 0.2 & 1 & \ 9.1 & 0.5 & 11 & 2.6 & 0.2 & \ 3 \\
H409    & 1.7 & 0.1 & 2 & 4.1 & 0.2 & 7 & 12.1 & 0.6 & 46 & 3.3 & 0.2 & 17 \\
H67     & \nodata & \nodata & \nodata & 4.4 & 0.4 & 6 & 11.2 & 0.7 & 38 & 2.6 & 0.4 & 11 \\
H70     & \nodata & \nodata & \nodata & 4.0 & 0.5 & 3 & 12.3 & 0.7 & 24 & 4.0 & 0.3 & 10  \\
H71     & 1.8 & 0.3 & 2 & 4.1 & 0.4 & 8 & 11.1 & 0.6 & 53 & 3.0  & 0.4 & 16  \\
H875    & \nodata & \nodata & \nodata & \nodata & \nodata & \nodata & \ 9.7 & 0.6 & 14 & \nodata & \nodata & \nodata \\
H972    & \nodata & \nodata & \nodata & \nodata & \nodata & \nodata & \ 9.7 & 0.7 & \ 8 & 2.5 & 0.5 & \ 3 \\
H974    & \nodata & \nodata & \nodata & \nodata & \nodata & \nodata & \ 7.6 & 2.9 & \ 1 & \nodata & \nodata & \nodata \\
N5471-A & 1.6 & 0.1 & 2 & 3.8 & 0.2 & 6 & 10.9 & 0.6 & 28 & 3.2 & 0.2 & 15 \\
N5471-B & 1.7 & 0.1 & 2 & 3.2 & 0.2 & 5 & 10.8 & 0.6 & 38 & 2.9 & 0.2 & 14 \\
N5471-C & 1.5 & 0.2 & 2 & 3.4 & 0.2 & 6 & 10.8 & 0.6 & 39 & 2.5 & 0.2 & 13 \\
N5471-D & 1.6 & 0.1 & 1 & 0.9 & 0.1 & 2 & 10.6 & 0.6 & 31 & 3.0 & 0.2 & 11 \\
SDH323  & \nodata & \nodata & \nodata & 2.9 & 0.7 & 2 & 15.2 & 1.1 & 35 & \nodata & \nodata & \nodata \\
\enddata
\tablenotetext{a}{Intensities are normalized to $I(H\beta)=100$, with
the corresponding uncertainty listed in the second column.}
\end{deluxetable}


\begin{deluxetable}{lcccccccccccc}
\tabletypesize{\scriptsize}
\tablecolumns{13}
\tablewidth{0pt}
\tablenum{8}
\tablecaption{He$^+$ and Total He Abundances}

\tablehead{
\colhead{ID}            &
\multicolumn{10}{c}{He$^+$/H$^+$}                      &
\multicolumn{2}{c}{He/H\tablenotemark{a}}                      \\
\colhead{  }            &
\colhead{$\lambda$4026}                      &
\colhead{err}                      &
\colhead{$\lambda$4471}                      &
\colhead{err}                      &
\colhead{$\lambda$5876}                      &
\colhead{err}                      &
\colhead{$\lambda$6678}                      &
\colhead{err}                      &
\colhead{avg.}                      &
\colhead{err}                      &
\colhead{  }                      &
\colhead{  }                      
}
\startdata
\vspace{-3mm}
\\
H1013   & 0.10  & 0.01  & 0.081 & 0.005 & 0.083 & 0.006 & 0.109 & 0.006 & 0.091 & 0.006 & 0.11  & 0.02  \\
H1105   & 0.10  & 0.01  & 0.087 & 0.005 & 0.091 & 0.006 & 0.093 & 0.006 & 0.091 & 0.006 & 0.099 & 0.008  \\
H1159   & 0.12  & 0.03  & 0.091 & 0.009 & 0.093 & 0.006 & 0.093 & 0.006 & 0.093 & 0.007 & 0.095 & 0.008  \\
H1170   & 0.08  & 0.01  & 0.076 & 0.007 & 0.078 & 0.006 & 0.074 & 0.006 & 0.077 & 0.007 & 0.092 & 0.01   \\
H1176   & 0.11  & 0.01  & 0.079 & 0.007 & 0.087 & 0.005 & 0.091 & 0.006 & 0.090 & 0.006 & 0.092 & 0.007  \\
H1216   & 0.09  & 0.01  & 0.082 & 0.005 & 0.082 & 0.005 & 0.092 & 0.006 & 0.086 & 0.006 & 0.087 & 0.007  \\
H128    & 0.09  & 0.01  & 0.082 & 0.005 & 0.092 & 0.006 & 0.090 & 0.006 & 0.088 & 0.006 & 0.090 & 0.007  \\
H143    & \nodata  & \nodata  & 0.08  & 0.01  & 0.096 & 0.006 & 0.097 & 0.009 & 0.094 & 0.007 & 0.095 & 0.008  \\
H149    & 0.10  & 0.01  & 0.085 & 0.007 & 0.093 & 0.006 & 0.093 & 0.006 & 0.092 & 0.007 & 0.092 & 0.008  \\
H336    & \nodata  & \nodata  & 0.057 & 0.005 & 0.064 & 0.005 & 0.067 & 0.006 & 0.063 & 0.005 & 0.11  & 0.04  \\
H409    & 0.089 & 0.005 & 0.086 & 0.005 & 0.089 & 0.006 & 0.087 & 0.007 & 0.088 & 0.006 & 0.088 & 0.007  \\
H67     & \nodata  & \nodata  & 0.094 & 0.009 & 0.085 & 0.007 & 0.07  & 0.01  & 0.086 & 0.008 & 0.09  & 0.01   \\
H70     & \nodata  & \nodata  & 0.092 & 0.012 & 0.094 & 0.005 & 0.11  & 0.01  & 0.097 & 0.006 & 0.099 & 0.007  \\
H71     & \nodata  & \nodata  & 0.086 & 0.008 & 0.085 & 0.005 & 0.084 & 0.011 & 0.085 & 0.006 & 0.085 & 0.006  \\
N5471-A & 0.087 & 0.006 & 0.082 & 0.005 & 0.085 & 0.005 & 0.090 & 0.007 & 0.085 & 0.006 & 0.085 & 0.006  \\
N5471-B & 0.092 & 0.006 & 0.076 & 0.005 & 0.086 & 0.005 & 0.083 & 0.006 & 0.084 & 0.006 & 0.084 & 0.007  \\
N5471-C & 0.081 & 0.006 & 0.073 & 0.005 & 0.083 & 0.005 & 0.070 & 0.007 & 0.078 & 0.006 & 0.078 & 0.007  \\
N5471-D & 0.086 & 0.006 & 0.019 & 0.005 & 0.082 & 0.005 & 0.084 & 0.006 & 0.082 & 0.006 & 0.082 & 0.007  \\
SDH323  & \nodata  & \nodata  & 0.07  & 0.01  & 0.12  & 0.01  & \nodata  & \nodata & 0.07  & 0.01  & 0.09  & 0.02  \\
H681\tablenotemark{b}  & \nodata  & \nodata & 0.07 &  0.01 & 0.06  & 0.01 & 0.06 & 0.03 &  0.07 & 0.01 & 0.074 & 0.01  \\
\enddata
\tablenotetext{a}{Corrected for neutral He as described in text.}
\tablenotetext{b}{From Garnett \& Kennicutt 1994.}
\end{deluxetable}

\end{document}